\documentclass[preprint,12pt]{elsarticle}

\usepackage{amsmath,amsfonts,amssymb}
\usepackage{algorithmic}
\usepackage{algorithm}
\usepackage{array}
\usepackage{textcomp}
\usepackage{stfloats}
\usepackage{url}
\usepackage{verbatim}
\usepackage{graphicx}
\usepackage[numbers]{natbib}
\usepackage{siunitx}
\usepackage{soul}
\usepackage{booktabs}
\usepackage{tablefootnote}
\usepackage{threeparttable}
\usepackage{multicol}
\usepackage{multirow}
\usepackage{bm}
\usepackage{hyperref}
\usepackage{enumitem}
\usepackage{caption}
\usepackage[labelformat=empty]{subcaption}
\usepackage{makecell}
\usepackage{pifont}
\usepackage{wrapfig}
\usepackage{setspace}

\hypersetup{hidelinks=true}
\newcommand{\etal}{\textit{et~al.}}
\newcommand{\CustomComment}[1]{\(\hfill \triangleright\) \textit{#1}}

\DeclareMathOperator{\atantwo}{atan2}
\newcommand{\cmark}{\ding{51}}  
\newcommand{\xmark}{\ding{55}}  

\journal{Computers in Biology and Medicine}

\begin{document}

\begin{frontmatter}


\author{Blake VanBerlo\corref{cor1}\fnref{uwcs}}
\ead{bvanberl@uwaterloo.ca}
\author{Jesse Hoey\fnref{uwcs}}
\author{Alexander Wong\fnref{uwsde}}
\author{Robert Arntfield\fnref{uwomed}}

\title{The Efficacy of Semantics-Preserving Transformations in Self-Supervised Learning for Medical Ultrasound}

\affiliation[uwcs]{organization={Cheriton School of Computer Science, University of Waterloo},
            addressline={200 University Ave W},
            city={Waterloo},
            postcode={N2L 3G1},
            state={Ontario},
            country={Canada}}

\affiliation[uwsde]{organization={Department of Systems Design Engineering, University of Waterloo},
            addressline={200 University Ave W},
            city={Waterloo},
            postcode={N2L 3G1},
            state={Ontario},
            country={Canada}}

\affiliation[uwomed]{organization={Schulich School of Medicine and Dentistry, Western University},
            addressline={1151 Richmond St},
            city={London},
            postcode={N6A 3K7},
            state={Ontario},
            country={Canada}}


\begin{abstract}
Data augmentation is a central component of joint embedding self-supervised learning (SSL).
Approaches that work for natural images may not always be effective in medical imaging tasks.
This study systematically investigated the impact of data augmentation and preprocessing strategies in SSL for lung ultrasound.
Three data augmentation pipelines were assessed: (1) a baseline pipeline commonly used across imaging domains, (2) a novel semantic-preserving pipeline designed for ultrasound, and (3) a distilled set of the most effective transformations from both pipelines.
Pretrained models were evaluated on multiple classification tasks: B-line detection, pleural effusion detection, and COVID-19 classification.
Experiments revealed that semantics-preserving data augmentation resulted in the greatest performance for COVID-19 classification - a diagnostic task requiring global image context.
Cropping-based methods yielded the greatest performance on the B-line and pleural effusion object classification tasks, which require strong local pattern recognition.
Lastly, semantics-preserving ultrasound image preprocessing resulted in increased downstream performance for multiple tasks.
Guidance regarding data augmentation and preprocessing strategies was synthesized for developers working with SSL in ultrasound.
\end{abstract}


\begin{highlights}
\item Novel semantics-preserving data augmentation for self-supervision in ultrasound
\item Semantics-preserving augmentation yielded best results for COVID-19 classification
\item Cropping-based data augmentation was best for object classification tasks 
\item Semantics-preserving ultrasound preprocessing improves pretrained model performance
\item Guidance is provided for developers working with unlabeled ultrasound data
\end{highlights}


\end{frontmatter}



\section{Introduction}
\label{sec:introduction}

Automated interpretation of medical ultrasound (US) images is increasingly implemented using deep learning~\cite{wang2021deep}.
Deep neural networks (DNN) achieve strong performance for applications in US imaging, such as distinguishing benign from malignant liver lesions~\cite{yang2020improving}, estimating left ventricular end-diastolic and end-systolic volumes~\cite{ghorbani2020deep}, and screening for pneumothorax~\cite{vanberlo2022accurate}.

Despite early successes, investigators are limited by the lack of publicly available datasets~\cite{liu2019deep,ansari2024advancements}.
When available, researchers use private collections of US examinations, as they may contain far more samples.
Given the expense of manual annotation, many are turning to self-supervised learning (SSL) methods to pretrain DNNs using large, unlabeled collections of US data~\cite{vanberlo2024survey}.
These SSL-pretrained backbone DNNs may be fine-tuned for supervised learning tasks of interest.

An important category of SSL methods for computer vision is the \textit{joint embedding architecture}, which is characterized by training DNNs to produce similar vector representations for pairs of related images.
The most common method for retrieving related pairs of images from unlabeled datasets is to apply random transformations (i.e., data augmentation) to an image, producing two distorted views.
The choice of random transformations steers the invariance relationships learned by the backbone.

\begin{figure*}
    \centering
    \begin{subfigure}[b]{0.16\textwidth}
        \centering 
        \includegraphics[height=5cm]{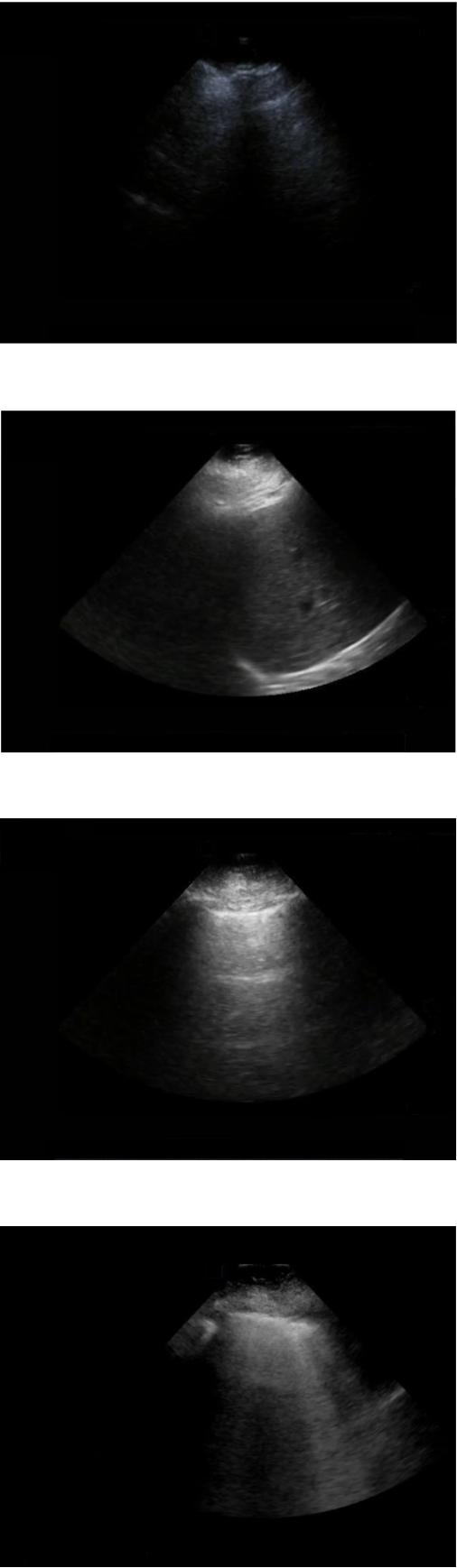}
        \caption{\textbf{(a)} Original \\ images}
    \end{subfigure}
    \hfill
    \begin{subfigure}[b]{0.16\textwidth}
        \centering 
        \includegraphics[height=5cm]{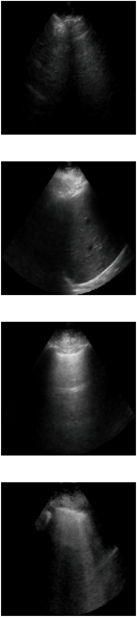}
        \caption{\textbf{(b)} Preprocessed images}
    \end{subfigure}
    \hfill
    \begin{subfigure}[b]{0.21\textwidth}
        \centering 
        \includegraphics[height=5cm]{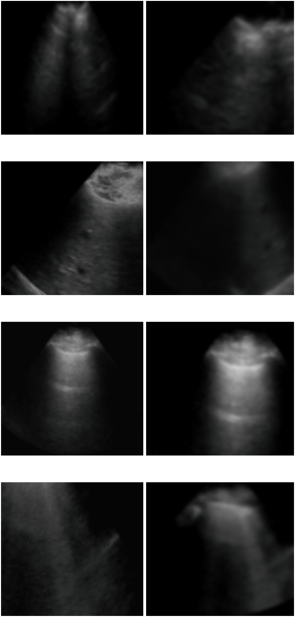}
        \caption{\textbf{(c)} StandardAug \\ pipeline}
    \end{subfigure}
    \hfill
    \begin{subfigure}[b]{0.21\textwidth}
        \centering 
        \includegraphics[height=5cm]{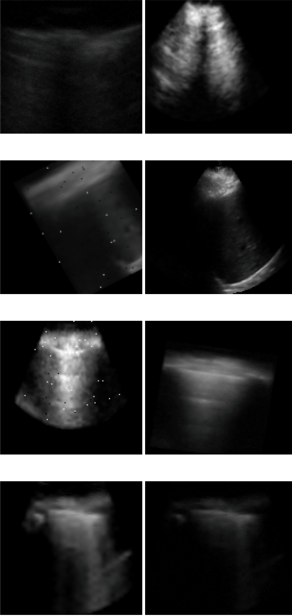}
        \caption{\textbf{(d)} AugUS-O \\ pipeline}
    \end{subfigure}
    \hfill
    \begin{subfigure}[b]{0.21\textwidth}
        \centering 
        \includegraphics[height=5cm]{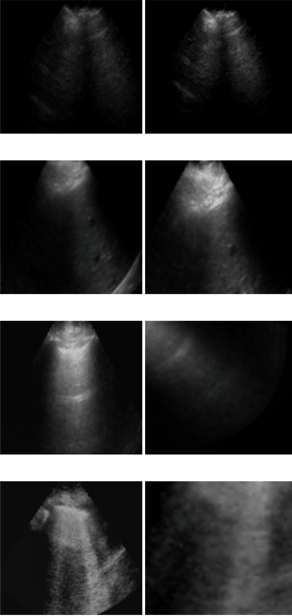}
        \caption{\textbf{(e)} AugUS-D \\ pipeline}
    \end{subfigure}
    \caption{Examples of the preprocessing and data augmentation methods in this study.
    \textbf{(a)} Original images are from ultrasound exams. \textbf{(b)} Semantics-preserving preprocessing is applied to crop out areas external to the field of view. \textbf{(c)} The StandardAug pipeline is a commonly employed data augmentation pipeline in self-supervised learning. \textbf{(d)} The AugUS-O pipeline was designed to preserve semantic content in ultrasound images. \textbf{(e)} AugUS-D is a hybrid pipeline whose construction was informed by empirical investigations into the StandardAug and AugUS-O pipelines. }
    \label{fig:overview}
\end{figure*}

In this study, we proposed and assessed data preprocessing and data augmentation strategies designed to preserve semantic content in medical ultrasound images (Fig~\ref{fig:overview}).
We compared handcrafted domain-specific augmentation methods against standard SSL data augmentation practises.
We found that US-specific transformations resulted in the greatest improvement in performance for COVID-19 classification -- a diagnostic task -- on a public dataset.
Experiments also revealed that standard cropping-based augmentation strategies outperformed US-specific transformations for object classification tasks in lung US. 
Lastly, ultrasound-specific semantics-preserving preprocessing was found to be instrumental to the success of pretrained backbones.
In summary, our contributions are as follows:
\begin{itemize}
    \item Semantics-preserving image preprocessing for SSL in US
    \item Semantics-preserving data augmentation methods designed for US images 
    \item Comparison of multiple data augmentation strategies for SSL for multiple types of lung US tasks
    \item Recommendations for developers working with unlabeled US datasets
\end{itemize}
To our knowledge, this study is the first to quantify the impact of data augmentation methods for SSL with US.
We are hopeful that the results and lessons from this study may contribute to the development of foundation models for medical US.

\section{Background}
\label{sec:background}

\subsection{Data Augmentation in Self-Supervised Learning}

The joint embedding class of SSL methods is characterized by the minimization of an objective function that, broadly speaking, encourages similarity of related pairs of inputs.
Semantically related pairs of images (i.e., \textit{positive pairs}) are sampled from unlabeled datasets according to a \textit{pairwise relationship}.
If the SSL pairwise relationship is satisfied for samples exhibiting the same class, SSL methods will likely improve the performance of a classifier~\cite{Balestriero2022}.
Most joint embedding methods rely on data augmentation to define the pairwise relationship.
Some studies have used meta-data or known relationships between samples to identify related pairs~\cite{azizi_big_2021,zhao2021longitudinal,basu2022unsupervised}; however, the availability of such information is rare.
The choice of data augmentation transformations is therefore crucial, as it dictates the invariances learned~\cite{cabannes2023ssl}.
However, the set of useful invariances differs by the image modality and downstream problem(s) of interest. 
Despite this, studies continue to espouse a data augmentation pipeline popularized by leading SSL methods, such as SimCLR~\cite{chen2020simple}, BYOL~\cite{grill2020bootstrap}, Barlow Twins~\cite{zbontar2021barlow}, and VICReg~\cite{bardes2022vicreg}.
These methods utilized the same pipeline, but with minor hyperparameter variations.
The pipeline includes the following transformations: random crops, horizontal reflection, colour jitter, Gaussian blur, and solarization.
Hereafter, we refer to this baseline pipeline as \textit{StandardAug}.
Random rotation is an example of a transformation not found in the StandardAug pipeline that represents an important invariance relationship for many tasks in medical imaging.
For example, random rotation has been applied in SSL pretraining with magnetic resonance exams of the prostate~\cite{fernandez2022contrasting}.
Moreover, the authors did not use StandardAug's Gaussian blur transformation because it may have rendered the images uninterpretable.

\subsection{Joint Embedding Self-Supervision in Ultrasound}

Recent studies have examined the use of joint embedding SSL methods for US interpretation tasks, such as echocardiogram view classification~\cite{anand_benchmarking_2022}, left ventricle segmentation~\cite{saeed_contrastive_2022}, and breast tumor classification~\cite{nguyen2021semi}. 
Some have proposed positive pair sampling schemes customized for US.
The USCL method and its successors explored contrastive learning methods where the positive pairs were weighted sums of images from the same US video~\cite{chen2021uscl,chen2022generating,zhang2022hico}.
Other methods have studied the use of images from the same video as positive pairs~\cite{basu2022unsupervised,vanberlo2024intra}.
In these studies, the set of transformations were a subset of the StandardAug data augmentation pipeline, occasionally with different hyperparameters.
Few studies have proposed US-specific data augmentation methods for SSL.
A recent study by Chen~\textit{et al.}~\cite{chen2023contrastive} applied BYOL and SimCLR to pretrain 3D convolutional DNNs with specialized data augmentation for lung consolidation detection in US videos, observing that temporal transformations were contributory to their problem.
This study builds on previous literature by proposing and comparing domain-specific data augmentation and preprocessing method for multiple types of downstream tasks.

\section{Materials \& Methods}
\label{sec:method}

\subsection{Datasets and Tasks}
\label{subsec:datasets}

\begin{table}[h!]
    \caption{Breakdown of the unlabelled, training, validation, and test sets in the private dataset. 
    For each split, we indicate the number of distinct patients, videos, and images.
    $x / y$ indicates the number of labeled videos in the negative and positive class for each binary classification task.
    For the {\tt PL} task, we indicate the number of videos with frame-level bounding box annotations.}
    \label{tab:private-dataset}
    \centering
    \setlength{\tabcolsep}{0.17cm}
    \begin{tabular}{cccccc}
        \toprule
         & \multicolumn{4}{c}{Local} & External \\
         \cmidrule(r){2-5} \cmidrule(l){6-6}
         & Unlabeled & Train & Validation & Test & Test \\
         \midrule
         Patients & $5571$ & $1702$ & $364$ & $364$ & $168$ \\ 
         Videos & $59309$ & $5679$ & $1184$ & $1249$ & $925$ \\ 
         Images & \num{1.3e7} & \num{1.2e6} & \num{2.5e5} & \num{2.6e5} & \num{1.1e5} \\
         {\tt AB} labels & N/A & $2067 / 999$ & $459 / 178$ & $458 / 221$ & $286 / 327$ \\ 
         {\tt PE} labels & N/A &  $789 / 762$ & $176 / 142$ & $162 / 158$ & $68 / 110$ \\ 
         {\tt PL} labels & N/A &  $200$ & $39$ & $45$ & $0$ \\ 
         \bottomrule
    \end{tabular}
\end{table}

We assessed the methods in this publication using a combination of public and private data.
COVIDx-US is a public COVID-19 lung US dataset consisting of $242$ publicly sourced videos, acquired from a variety of manufacturers and sites~\cite{Ebadi2022-mn}.
Each example is annotated with one of the following classes: normal, COVID-19 pneumonia, non-COVID-19 pneumonia, and other lung pathology.
Referred to as {\tt COVID} hereafter, the task is a four-class image classification problem.
Since there is no standard test partition, we split the data by patient identifier into training (70\%), validation (15\%), and test (15\%) splits.

The second data source is a private collection of lung ultrasound examinations, and we refer to it as \textit{LUSData}.
Access to this data was granted by Western University research ethics board (REB 116838) on January 28, 2021.
LUSData contains videos from of parenchymal and pleural views of the lung.
A subset of the parenchymal views has labels for the presence of A-lines or B-lines (i.e., the {\tt AB} classification task).
A-lines are reverberation artifacts that indicate normal lung tissue, while B-lines are axial artifacts that indicate fluid or thickness in the lung.
A subset of the pleural views is labeled for the presence or absence of pleural effusion (i.e., the {\tt PE} classification task), which is an accumulation of fluid around the lungs.
A small fraction of the parenchymal views in LUSData possess bounding box labels for the pleural line  (i.e., the {\tt PL} object detection task).
Most exams in LUSData originated from a local healthcare centre, but a subset were acquired at another centre, which we adopt as an external test set.
The labeled examples in the local dataset were split into training (70\%), validation (15\%), and test (15\%) splits by patient.
Table~\ref{tab:private-dataset} provides the video and class counts of LUSData.
Further dataset details are in \ref{apx:dataset-details}.
All models in this study are trained on images, instead of on videos.
Classification labels apply to every image in the video.
However, individual images within each video labeled for the {\tt PL} task have bounding box annotations.

\subsection{Semantics-Preserving Preprocessing}
\label{subsec:semantic-preserving-preprocessing}

\begin{figure}
    \centering
    \includegraphics[width=\linewidth]{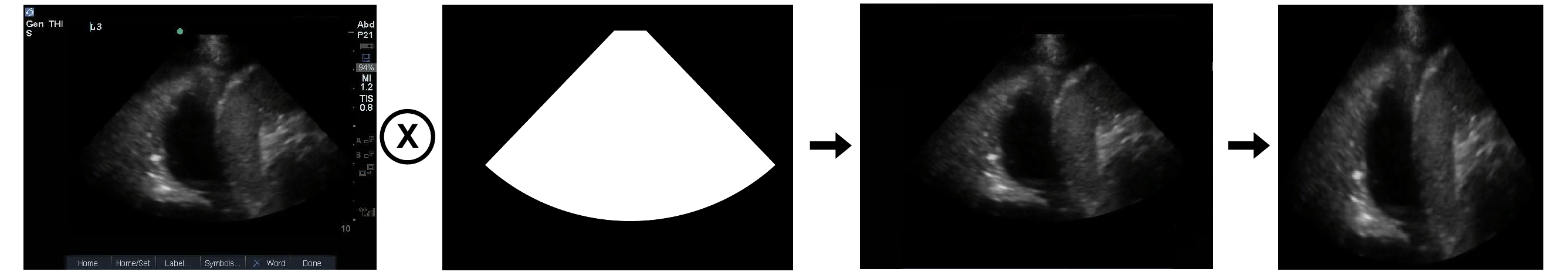}
    \caption{Raw ultrasound images are preprocessed by applying a mask that preserves only the field of view, then cropped according to the bounds of the field of view.}
    \label{fig:semantic-preprocessing}
\end{figure}

The field of view (FOV) in US images is typically surrounded by burnt-in scan parameters, logos, and other details.
We estimated the shape of the FOV and masked out all extraneous graphical entities using ultrasound cleaning software (UltraMask, Deep Breathe Inc., London, ON, Canada).
Semantic information only exists within the FOV of the US, which typically occupies a fraction of the images.
Scaling transformations, such as random cropping, could produce views that largely contain background.
Accordingly, we cropped the cleaned images to the smallest rectangle that encapsulates the FOV mask to maximize semantic content in US images.
Fig.~\ref{fig:semantic-preprocessing} depicts this semantics-preserving preprocessing workflow.
The process was applied to all images in LUSData and COVIDx-US.

\subsection{Ultrasound-specific Data Augmentation}
\label{subsec:us-augmentations}

Joint embedding SSL is effective when positive pairs contain similar information with respect to downstream tasks~\cite{Balestriero2022}.
Several SSL studies applied to photographic or medical imaging datasets adopted variations of the StandardAug data augmentation pipeline.
The core aim of our study was to determine if semantics-preserving data augmentation would better equip pretrained feature extractors for downstream LUS tasks than the commonly applied StandardAug pipeline.

We refer to a \textit{data augmentation pipeline} as an ordered sequence of transformations, each applied with some probability.
For clarity, we assign each transformation an alphanumeric identifier and express a data augmentation pipeline as an ordered sequence of identifiers.
The StandardAug pipeline transformations are detailed in Table~\ref{tab:byol-augs}.
The table also includes an estimate of the time to transform a single image.
Details on how the runtime estimates were calculated are in \ref{apx:transformation-runtime-estimates}.

\begin{table}[h!]
    \centering
    \caption{The sequence of transformations in the StandardAug data augmentation pipeline}
    \begin{tabular}{ccp{6cm}c}
        \toprule
        Identifier & Probability & Transformation & Time [ms] \\
        \midrule
        \textit{B00} & 1.0 & Crop and resize & $0.29$ \\
        \textit{B01} & 0.5 & Horizontal reflection & $0.08$ \\
        \textit{B02} & 0.8 & Color jitter. & $2.40$ \\
        \textit{B03} & 0.2 & Conversion to grayscale & $0.19$ \\
        \textit{B04} & 0.5 & Gaussian blur & $0.74$ \\
        \textit{B05} & 0.1 & Solarization & $0.15$ \\
        \bottomrule
    \end{tabular}
    \label{tab:byol-augs}
\end{table}

\begin{figure}
    \centerline{\includegraphics[width=\textwidth]{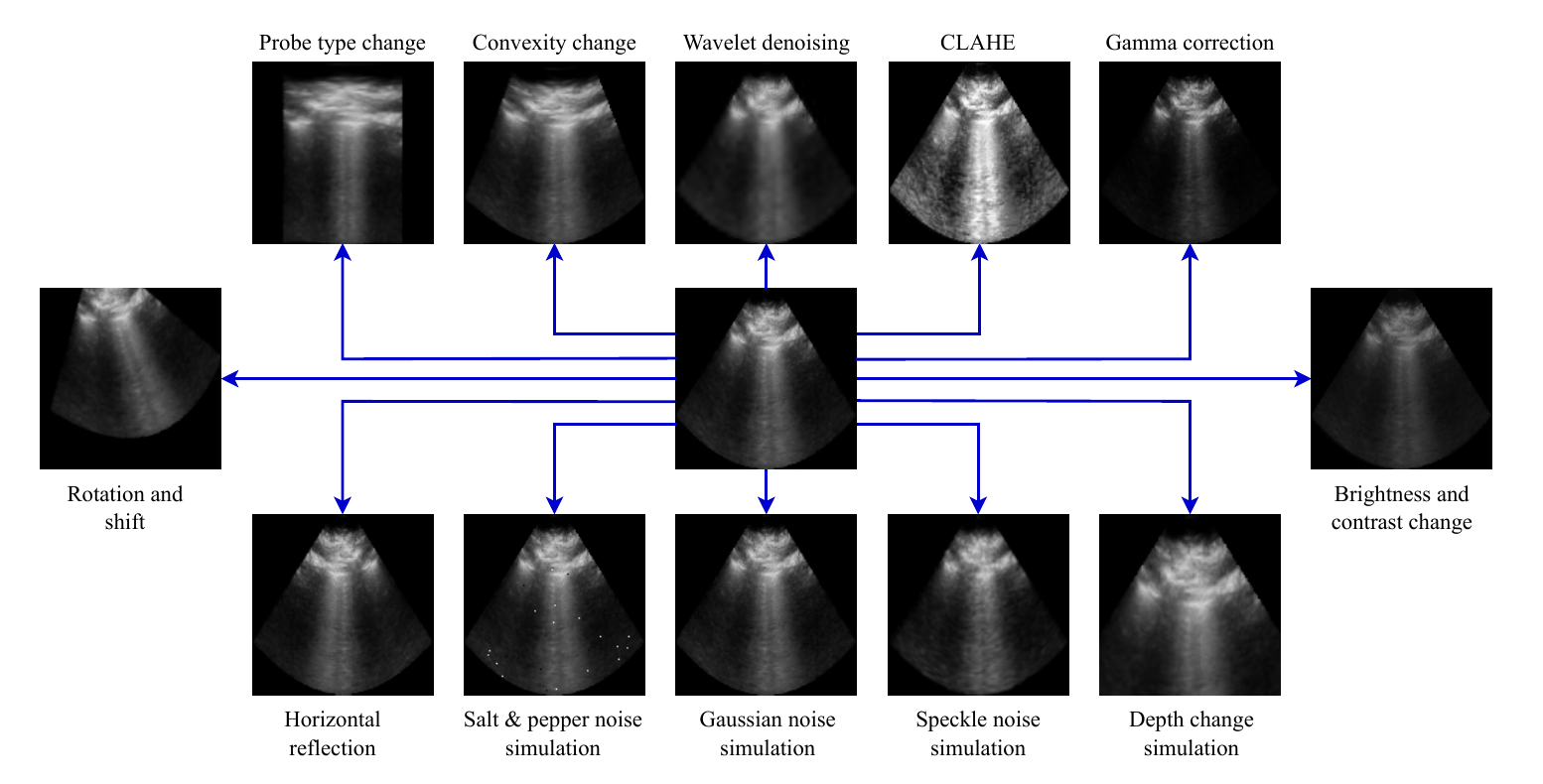}}
    \caption{Examples of ultrasound-specific data augmentation transformations applied to the same US image.}
    \label{fig:auguso-augmentations}
\end{figure}

We designed the \textit{AugUS-O} pipeline, which was intended to preserve semantic information in the entire US FOV while imposing nontrivial differences across invocations.
The transformations in AugUS-O are listed below.

\begin{enumerate}[label={}]
\item[B00:] \textbf{Probe Type Change:} 
Inspired by Zeng~\etal's~work~\cite{zeng2024covid}, this transformation resamples a US image according to a different field of view (FOV) shape. 
Linear FOV shapes are converted to curvilinear shapes, while curvilinear and phased array are converted to linear. 
\item[U01:] \textbf{Convexity Change:} 
The shape of convex FOVs can vary, depending on the manufacturer, depth, and field of view of the probe.
This transformation modifies the FOV shape such that the distance between $x_1$ and $x_2$ is altered, mimicking a change in $\theta_0$.
\item[U02:] \textbf{Wavelet Denoising: } 
As an alternative to the commonly used Gaussian blur transformation, this transformation denoises an image by thresholding it in wavelet space, according to Birg{\'e} and Massart's method~\cite{birge1997model}.
\item[U03:] \textbf{Contrast-Limited Adaptive Histogram Equalization: }
This transformation enhances contrast by applying locally informed equalization~\cite{pizer1987}. 
\item[U04:] \textbf{Gamma Correction: }
In contrast to standard brightness change transforms, gamma correction applies a nonlinear change in pixel intensity.
\item[U05:] \textbf{Brightness and Contrast Change: }
The brightness and contrast of the image are modified by applying a linear transform to the pixel values.
\item[U06:] \textbf{Depth Change Simulation: } 
Changing the depth controls on an ultrasound probe impacts how far the range of visibility is from the probe.
This transformation simulates a change in depth by applying a random zoom, while preserving the FOV shape. 
\item[U07:] \textbf{Speckle Noise Simulation:} 
Speckle noise, Gaussian noise, and salt \& pepper (S\&P) noise are prevalent in US~\cite{vilimek2022comparative}. 
This transformation applies Singh~\etal's~\cite{singh2017synthetic} synthetic speckle noise algorithm to the image. 
\item[U08:] \textbf{Gaussian Noise Simulation:}
Multiplicative Gaussian noise is independently applied to each pixel.
\item[U09:] \textbf{Salt \& Pepper Noise Simulation: } 
A small, random assortment of pixels are set to black or white.
\item[U10:] \textbf{Horizontal Reflection: } 
The image is reflected about the central vertical.
\item[U11:] \textbf{Rotation \& Shift: } 
The image is rotated and translated by a random angle and vector, respectively.
\end{enumerate}

Refer to Figure~\ref{fig:auguso-augmentations} for a visual example of each transformation in AugUS-O.
Algorithmic details and parameter settings for the StandardAug and AugUS-O pipelines are in Appendices~\ref{apx:byol-transformations}~and~\ref{apx:us-specific-transformations}, respectively.
As is common in stochastic data augmentation, each transformation was applied with some probability.
Table~\ref{tab:us-specific-augs} gives the entire sequence of transformations, the probability with which each is applied.
Visuals of positive pairs produced using the StandardAug and AugUS-O augmentation pipelines can be found in Fig.~\ref{subfig:aug-examples-1} and Fig.~\ref{subfig:aug-examples-2}, respectively.

\begin{table}[h!]
    \centering
    \caption{The sequence of transformations in the US-specific augmentation pipeline.}
    \setlength{\tabcolsep}{0.06cm}
    \begin{threeparttable}
    \begin{tabular}{ccp{6cm}c}
        \toprule
        Identifier & Probability & Transformation & Time [ms] \\
        \midrule
        U00 & 0.3 & Probe type change & $2.25$ \\
        U01 & 0.75 & Convexity change  & $1.92$ \\
        U02 & 0.5 & Wavelet denoising & $5.00$ \\
        U03 & 0.2 & CLAHE\tnote{\textdagger} & $4.64$ \\
        U04 & 0.5 & Gamma correction & $0.52$ \\
        U05 & 0.5 & Brightness and contrast change & $0.49$ \\
        U06 & 0.5 & Depth change simulation & $1.76$ \\
        U07 & 0.333 & Speckle noise simulation & $3.69$\\
        U08 & 0.333 & Gaussian noise & $0.28$ \\
        U09 & 0.1 & Salt \& pepper noise  & $0.18$ \\
        U10 & 0.5 & Horizontal reflection & $0.19$ \\
        U11 & 0.5 & Rotation \& shift & $1.42$ \\
        \bottomrule
    \end{tabular}
    \begin{tablenotes}
    \footnotesize
    \item[\textdagger] Contrast-limited adaptive histogram equalization
    \end{tablenotes}
    \end{threeparttable}
    \label{tab:us-specific-augs}
\end{table}

We conducted an informal assessment of the similarity of positive pairs.
Positive pairs were produced for $50$ randomly sampled images, using both the StandardAug and the AugUS-O pipelines.
The pairs were presented in random order to one of the authors, who is an expert in point-of-care US.
They were aware of the two pipelines but were not told which pipeline produced each pair.
The expert was asked to mark the pairs they believed conveyed the same clinical impression.
We observed that $58\%$ of pairs produced with the StandardAug pipeline were marked as similar, whereas $70\%$ of the AugUS-O pairs were marked as similar.
While not conclusive, this manual evaluation added credence to the semantics-preserving intention of the design.

\subsection{Discovering Semantically Contributory Transformations}

A major aim of this work was to explore the utility of various data augmentation schemes during pretraining.
As such, we conducted leave-one-out analysis for each of the StandardAug and AugUS pipelines to estimate the impact of each transformation on models' ability to solve downstream classification tasks.
We pretrained separate models on the unlabeled portion of LUSData, using an altered version of a pipeline with one transformation omitted.
We then conducted $10$-fold cross-validation on the LUSData training set for downstream classification tasks for each pretrained model.
The median cross-validation test performance for each model pretrained using an ablated pipeline was compared to a baseline model that was pretrained with the entire pipeline.
The experiment was conducted for both the StandardAug and AugUS pipelines.
Any transformations that, when omitted, resulted in worsened performance on either {\tt AB} or {\tt PE} were deemed contributory.

\subsection{Training Protocols}
\label{subsec:training-details}

We adopted the MobileNetV3Small architecture~\cite{howard2019searching} for all experiments in this study and pretrained using the SimCLR method~\cite{chen2020simple}.
MobileNetV3Small was chosen due to its real-time inference capability on mobile devices and its use in prior work by VanBerlo~\etal\, for similar tasks~\cite{vanberlo2023self}. 
The SimCLR projector was a $2$-layer multilayer perceptron with $576$ nodes per layer.
Images were resized to $128\times128$ pixels prior to the forward pass.
Unless otherwise stated, backbones (i.e., feature extractors) were initialized using ImageNet-pretrained weights~\cite{deng2009} and were pretrained using the LARS optimizer~\cite{you2019large} with a batch size of $1024$, a base learning rate of $0.2$ and a linear warmup with cosine decay schedule.
Pretraining was conducted for $3$ epochs with $0.3$ warmup epochs for LUSData, and $100$ epochs with $10$ warmup epochs for COVIDx-US.

To conduct supervised evaluation, a perceptron classification head was appended to the final pooling layer of the backbone.
Classifiers were trained using stochastic gradient descent with a momentum of $0.9$ and a batch size of $512$.
The learning rates for the backbone and head were $0.002$ and $0.02$, respectively; each was annealed according to a cosine decay schedule.
Training was conducted for $10$ epochs on LUSData and $30$ epochs on COVIDx-US.
Unless otherwise stated, the weights corresponding to the epoch with the lowest validation loss were retained for test set evaluation.

Although this study focused on classification tasks, we also evaluated backbones on the {\tt PL} object detection task using the Single Shot Detector (SSD) method~\cite{liu2016ssd}.
SSL-pretrained backbones were used as the convolutional feature extractor.
Architectural and training details for SSD are in \ref{apx:pl-od-training}.

Code for the experiments and transformations will be shared in a public GitHub repository upon publication.

\section{Results}
\label{sec:results}

\subsection{Transformation Leave-one-out Analysis}
\label{subsec:pipeline-design}

A leave-one-out analysis was conducted to discover which transformations in each of the StandardAug and AugUS-O pipelines were contributory to downstream task performance. 
We pretrained backbones using versions of each pipeline with one transformation omitted. 
The private LUSData training set was split by patient into $10$ disjoint subsets. 
For each pretrained backbone, $10$-fold cross-validation was conducted to obtain estimates of the performance of linear classifiers trained on its output feature vectors.
The maximum validation area under the receiver operating curve (AUC) across epochs was recorded.
Omitted transformations that resulted in statistically significant lower validation AUC for either the {\tt AB} or {\tt PE} task were included in a third pipeline.

We conducted statistical testing to compare each of the StandardAug and AugUS-O pipelines, and for each of the {\tt AB} and {\tt PE} tasks (described in Section~\ref{subsec:datasets}). 
Friedman's Test for multiple comparisons~\cite{friedman1940comparison} was conducted, with significance level $0.05$.
When significant differences were found, we performed the Wilcoxon Sign-Rank Test~\cite{wilcoxon1945} to compare the test AUCs from each ablated model to the baseline's test AUCs.
To control for false positives, the Holm-Bonferroni correction~\cite{holm1979simple} was applied to keep the family-wise significance level at $0.05$.

Table~\ref{tab:ablation-val-auc} details the results of the leave-one-out analysis.
Friedman's Test detected differences in performance on both the {\tt AB} and {\tt PE} tasks when pretrained using the StandardAug pipeline, but only the {\tt AB} task exhibited differences when pretrained with the AugUS-O pipeline.
As shown in Table~\ref{tab:ablation-val-auc}, the set of transformations that exhibited statistically significant reductions in test AUC for at least one task when excluded were crop \& resize (B00), color jitter (B02), CLAHE (U03), and rotation \& shift (U11).
\ref{apx:looa-stats} provides all test statistics from this investigation.
The random crop and resize transformation (B00) Of note is the sharp decrease in performance without the random crop and resize (B00), indicating that it is a critical transformation.

\begin{figure}
    \centering
    \subfloat[\textbf{(a)} StandardAug pipeline\label{subfig:aug-examples-1}]{%
    \includegraphics[width=0.7\columnwidth]{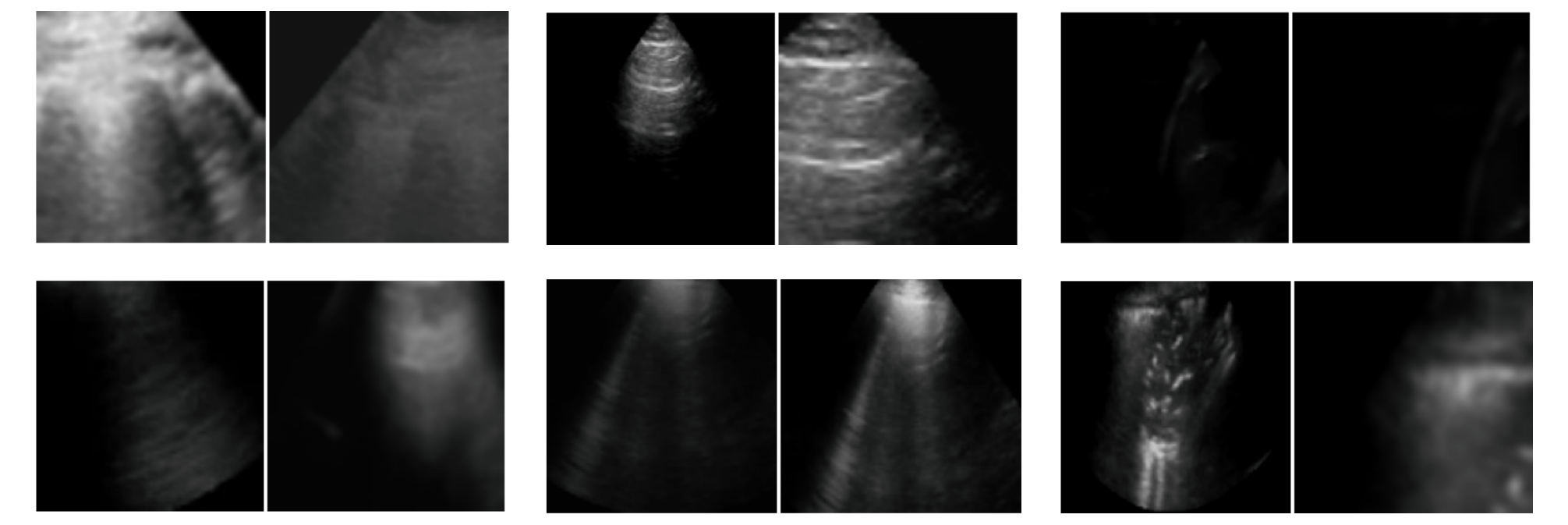}
    }
    \hfill
    \subfloat[\textbf{(b)} AugUS-O pipeline\label{subfig:aug-examples-2}]{%
    \includegraphics[width=0.7\columnwidth]{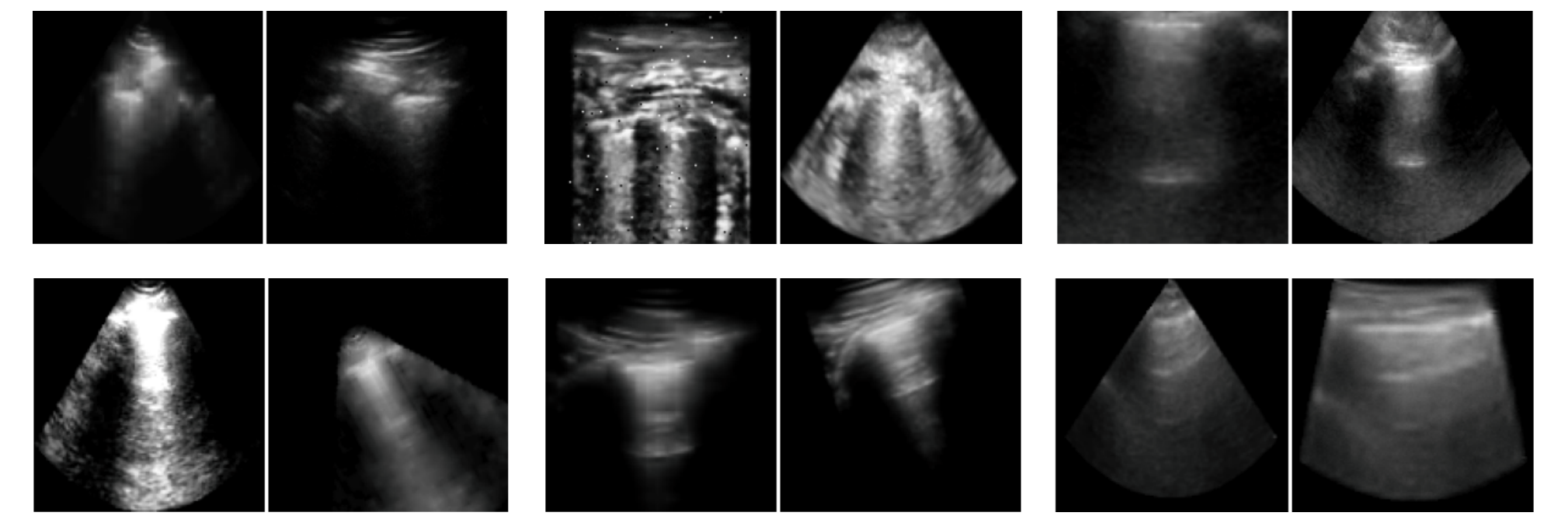}
    }
    \hfill
    \subfloat[\textbf{(c)} AugUS-D pipeline\label{subfig:aug-examples-3}]{%
    \includegraphics[width=0.7\columnwidth]{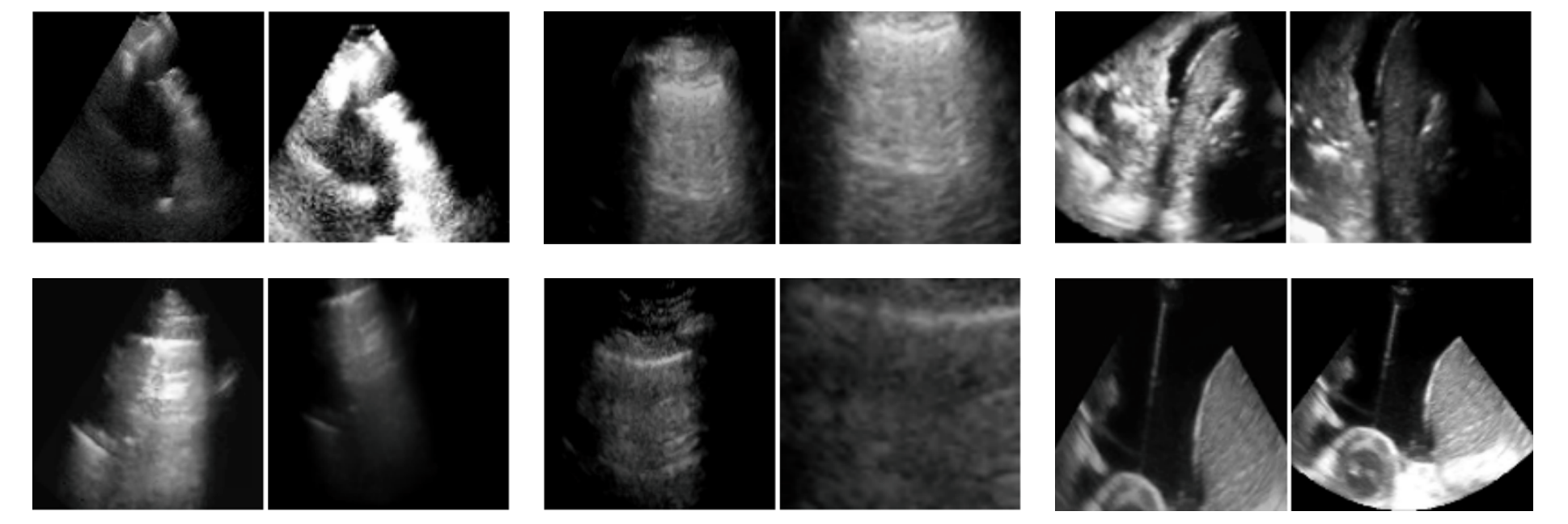}
    }
    \caption{Examples of positive pairs produced using each of the StandardAug, AugUS-O, and AugUS-D data augmentation pipelines.}
    \label{fig:aug-examples}
\end{figure}

Using these transformations, we constructed a distilled pipeline that consists only of the above transformations.
Referred hereafter to as \textit{AugUS-D}, the pipeline is expressed as the following sequence: \textit{[U03, B02, U11, B00]}.
Fig.~\ref{subfig:aug-examples-3} provides some examples of positive pairs produced with AugUS-D.
For more examples of pairs produced by each pipeline, see \ref{apx:additional-positive-pairs}.

\begin{table}[h!]
    \centering
    \caption{A comparison of ablated versions of the StandardAug and AugUS-O pipeline with one excluded transformation versus the original pipelines.
    Models were pretrained on the LUSData unlabeled set and evaluated on two downstream classification tasks -- {\tt AB} and {\tt PE}.
    Performance is expressed as mean and median test AUC from $10$-fold cross-validation achieved by a linear classifier trained on the feature vectors of a frozen backbone. 
    }
    \begin{threeparttable}
    \begin{tabular}{cccccc}
        \toprule
        Pipeline & Omitted  & \multicolumn{2}{c}{{\tt AB}} & \multicolumn{2}{c}{{\tt PE}}\\
         &  & Mean (std) & Median & Mean & Median \\
        \midrule
        \multirow{7}{*}{StandardAug} & None &  $0.978\,(0.007)$ & $0.978$ & $0.852\,(0.040)$ & $0.845$ \\
        \addlinespace[-0.5ex] \cmidrule{2-6} \addlinespace[-0.4ex]
        & {\it B00} & $0.864\,(0.022)$ & $0.873$\tnote{\textdagger} & $0.695\,(0.050)$& $0.707$\tnote{\textdagger}\\
        & {\it B01} &  $0.976\,(0.006)$ & $0.974$ & $0.848\,(0.046)$ & $0.856$\\
        & {\it B02} &  $0.975\,(0.007)$ & $0.975$\tnote{\textdagger} & $0.840\,(0.046)$ & $0.842$\tnote{\textdagger}\\
        & {\it B03} &  $0.978\,(0.007)$ & $0.978$ & $0.849\,(0.044)$ & $0.846$\\
        & {\it B04} &  $0.976\,(0.007)$ & $0.975$ & $0.840\,(0.041)$ & $0.842$\\
        & {\it B05} &  $0.977\,(0.007)$ & $0.977$ & $0.851\,(0.041)$ & $0.853$\\
        \midrule
        \multirow{13}{*}{AugUS-O} & None &  $0.956\,(0.013)$ & $0.959$ & $0.828\,(0.030)$ & $0.837$ \\
        \addlinespace[-0.5ex] \cmidrule{2-6} \addlinespace[-0.4ex]
        & {\it U00}  & $0.958\,(0.011)$ & $0.957$ & $0.831\,(0.034)$ & $0.839$\\
        & {\it U01}  & $0.952\,(0.016)$ & $0.952$ & $0.835\,(0.027)$ & $0.838$\\
        & {\it U02}  & $0.965\,(0.011)$ & $0.967$\tnote{\textsection} & $0.840\,(0.032)$ & $0.851$\\
        & {\it U03}  & $0.950\,(0.011)$ & $0.951$\tnote{\textdagger} & $0.825\,(0.028)$ & $0.827$\\
        & {\it U04}  & $0.957\,(0.013)$ & $0.958$ & $0.831\,(0.034)$ & $0.836$\\
        & {\it U05} & $0.953\,(0.014)$ & $0.952$ & $0.839\,(0.024)$ & $0.845$\\
        & {\it U06} & $0.961\,(0.009)$ & $0.959$ & $0.829\,(0.037)$ & $0.833$\\
        & {\it U07} & $0.959\,(0.012)$ & $0.960$& $0.838\,(0.035)$ & $0.856$\\
        & {\it U08}  & $0.961\,(0.013)$ & $0.966$\tnote{\textsection} & $0.834\,(0.027)$ & $0.849$\\
        & {\it U09}  & $0.962\,(0.012)$ & $0.967$\tnote{\textsection} & $0.838\,(0.030)$ & $0.845$\\
        & {\it U10}  & $0.956\,(0.011)$ & $0.959$ & $0.826\,(0.035)$ & $0.838$\\
        & {\it U11}  & $0.937\,(0.020)$ & $0.939$\tnote{\textdagger} & $0.825\,(0.028)$ & $0.823$\\
        \bottomrule
    \end{tabular}
    \begin{tablenotes}
    \footnotesize
    \item[\textdagger] Median is significantly less than baseline, where no transformations were omitted.
    \item[\textsection] Median is significantly greater than baseline, where no transformations were omitted.
    \end{tablenotes}
    \end{threeparttable}
    \label{tab:ablation-val-auc}
\end{table}

\subsection{Object Classification Task Evaluation}
\label{subsec:object-classification}

The StandardAug, AugUS-O, and AugUS-D pipelines were compared in terms of their performance on multiple downstream tasks.
Model backbones were pretrained using each of the data augmentation pipelines on the union of the unlabeled and training sets in LUSData.
Linear evaluation and fine-tuning experiments were performed according to the procedure explained in Section~\ref{subsec:training-details}.
In this section, we present results on the two object classification tasks: A-line vs B-line classification ({\tt AB}) and pleural effusion classification ({\tt PE}).

\begin{figure}[h!]
    \centering
    \includegraphics[width=0.9\columnwidth]{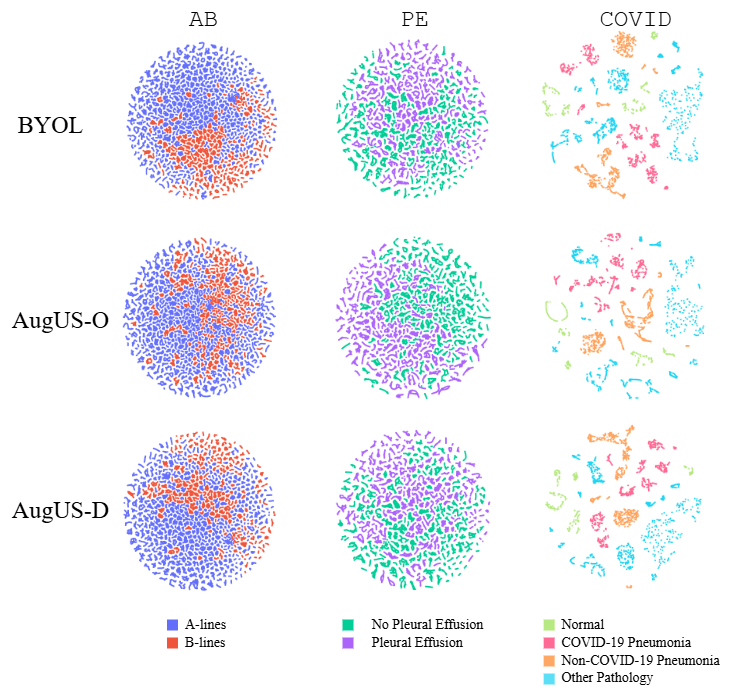}
    \caption{Two-dimensional t-SNE projections for test set feature vectors produced by SimCLR-pretrained backbones, for all tasks and data augmentation pipelines.}
    \label{fig:tsne}
\end{figure}

Linear classifiers indicate the usefulness of pretrained backbones, as the only trainable weights for supervised learning are those belonging to the perceptron head.
Table~\ref{tab:augus-internal-test-results} reports the test set performance of linear classifiers for each task and data augmentation pipeline.
On the private dataset, the AugUS-D and StandardAug pipelines performed comparably well on the {\tt AB} task.
AugUS-D attained greater performance metrics than the StandardAug pipeline on {\tt PE}.
To provide a visual perspective on linear classifier performance, we produced two-dimensional t-SNE embeddings of the feature vectors outputted by pretrained backbones.
Shown in Fig.~\ref{fig:tsne}, the separability of the visual representations is consistent with linear classifier performance.

\begin{table}[h!]
    \centering
    \caption{Test set performance for linear classification (LC) and fine-tuning (FT) experiments with the {\tt AB} and {\tt PE} tasks. 
    Binary metrics are averages across classes.
    The best observed metrics in each experimental setting are in \textbf{boldface}.}
    \setlength{\tabcolsep}{0.16cm}
    {\small
    \begin{tabular}{cccccccc}
        \toprule
         \makecell{Train \\ Setting} & Task & \makecell{Initial \\ Weights} & Pipeline & Accuracy & Precision & Recall & AUC \\
         \midrule 
          \multirow{8}{*}{{\large LC}} & \multirow{4}{*}{{\tt \large AB}} & SimCLR & StandardAug & $\mathbf{0.932}$ & $\mathbf{0.951}$ & $0.819$ & $0.970$ \\
          & & SimCLR & AugUS-O & $0.910$ & $0.939$ & $0.756$ & $0.953$ \\
          & & SimCLR & AugUS-D & $0.931$ & $0.947$ & $\mathbf{0.820}$ & $\mathbf{0.971}$ \\
          & & ImageNet & - & $0.898$ & $0.894$ & $0.758$ & $0.949$\\
          \cmidrule(l{3pt}){2-8} 
          & \multirow{4}{*}{{\tt \large PE}} & SimCLR & StandardAug & $0.782$ & $0.769$ & $0.787$ & $0.881$ \\
          & & SimCLR & AugUS-O & $0.795$ & $0.796$ & $0.776$ & $0.853$ \\
          & & SimCLR & AugUS-D & $\mathbf{0.800}$ & $\mathbf{0.798}$ & $0.782$ & $\mathbf{0.886}$ \\
          & & ImageNet & - & $0.779$ & $0.756$ & $\mathbf{0.804}$ & $0.864$ \\
         \midrule
           
          \multirow{10}{*}{{\large FT}} & \multirow{5}{*}{{\tt \large AB}} & SimCLR & StandardAug & $\mathbf{0.941}$ & $0.951$ & $0.850$ & $\mathbf{0.970}$\\
          & & SimCLR & AugUS-O & $0.939$ & $0.938$ & $\mathbf{0.859}$ & $0.968$\\
          & & SimCLR & AugUS-D & $0.931$ & $\mathbf{0.960}$ & $0.809$ & $0.962$\\
          & & Random & - & $0.883$ & $0.794$ & $0.834$ & $0.938$\\
          & & ImageNet & - & $0.911$ & $0.872$ & $0.830$ & $0.953$\\
          \cmidrule(l{3pt}){2-8} 
          & \multirow{5}{*}{{\tt \large PE}} & SimCLR & StandardAug & $0.766$ & $0.713$ & $0.863$ & $0.882$\\
          & & SimCLR & AugUS-O & $0.487$ & $0.479$ & $0.685$ & $0.557$\\
          & & SimCLR & AugUS-D & $\mathbf{0.802}$ & $0.782$ & $0.818$ & $\mathbf{0.884}$\\
          & & Random & - & $0.703$ & $\mathbf{0.733}$ & $0.607$ & $0.767$\\
          & & ImageNet & - & $0.708$ & $0.640$ & $\mathbf{0.907}$ & $0.845$\\
         \bottomrule
    \end{tabular}
    }
    \label{tab:augus-internal-test-results}
\end{table}

We fine-tuned the pretrained models, allowing the backbone's weights to be trainable in addition to the model head.
Table~\ref{tab:augus-internal-test-results} gives the test set performance of the fine-tuned classifiers.
We observed similar performance differences among the different augmentation pipelines, but note some additional findings.
The model pretrained using AugUS-O on LUSData performed comparably against the other pipelines on {\tt AB} but exhibited extremely poor performance on the {\tt PE} test set.
Although it may appear that this model may have overfit to the training set, examination of training metrics revealed that training and validation metrics were close, with validation set AUC having been evaluated as $0.861$.
Nonetheless, fine-tuned models that were pretrained with the StandardAug and AugUS-O pipelines yielded strong performance on both tasks in LUSData.

Fine-tuned classifiers for the {\tt AB} and {\tt PE} tasks were also evaluated on the external portion of the LUSData test set.
Most classifiers exhibited degraded performance on external data, compared to the local test set.
Overall, the relative performance of the classifiers on the external test set was reflective of their performance on local test data.
Unlike the local test evaluation, the network trained from scratch performed comparably to the SimCLR-pretrained models that utilized the AugUS-O and AugUS-D pipelines; however, the SimCLR-pretrained model that utilized the StandardAug pipeline achieved the greatest {\tt AB} test AUC by a margin of $0.029$.
On the {\tt PE} task, the classifier originating from the same pretrained model that utilized StandardAug achieved the greatest AUC by a margin of $0.015$.
Similar to the local test set, the pretrained models that incorporated the StandardAug and AugUS-D pipeline achieved the greatest test AUC, while the pretrained model that utilized AugUS-O performed the worst.

Although MobileNetV3Small was the backbone architecture used for these experiments, we repeated the above evaluations using the more commonly employed ResNet18 architecture~\cite{he2016deep}.
Similar trends were observed regarding the greater test performance attained by models pretrained with cropping-based pipelines.
However, the fine-tuned models greatly overfit, likely due to ResNet18's much greater capacity.
The ResNet18 models that achieved the greatest test performance were the linear classifiers trained on frozen backbones.
Notably, the trend persisted when evaluating on external test data.
Detailed results for ResNet18 can be found in \ref{apx:resnet18-experiments}.

\begin{table}[h!]
    \centering
    \caption{External test set metrics for linear classification (LC) and fine-tuning (FT) experiments with the {\tt AB} and {\tt PE} tasks. 
    Binary metrics are averages across classes.
    The best observed metrics in each experimental setting are in \textbf{boldface}.}
    \setlength{\tabcolsep}{0.13cm}
    {\small
    \begin{tabular}{cccccccc}
        \toprule
         \makecell{Train \\ Setting} & Task & \makecell{Initial \\ Weights} & Pipeline & Accuracy & Precision & Recall & AUC \\
         \midrule
         \multirow{8}{*}{{\large LC}} & \multirow{4}{*}{{\tt \large AB}} & SimCLR & StandardAug & $\mathbf{0.749}$ & $\mathbf{0.956}$ & $\mathbf{0.555}$ & $\mathbf{0.868}$  \\
          & & SimCLR & AugUS-O & $0.689$ & $0.927$ & $0.453$ & $0.810$  \\
          & & SimCLR & AugUS-D & $0.726$ & $0.922$ & $0.531$ & $0.859$  \\
          & & ImageNet & - & $0.643$ & $0.872$ & $0.389$ & $0.770$  \\
          \cmidrule(l{3pt}){2-8} 
          & \multirow{4}{*}{{\tt \large PE}} & SimCLR & StandardAug & $0.794$ & $0.835$ & $\mathbf{0.843}$ & $0.880$ \\
          & & SimCLR & AugUS-O & $0.784$ & $\mathbf{0.916}$ & $0.728$ & $0.870$ \\
          & & SimCLR & AugUS-D & $\mathbf{0.806}$ & $0.877$ & $0.809$ & $\mathbf{0.887}$ \\ 
          & & ImageNet & - & $0.758$ & $0.836$ & $0.771$ & $0.840$ \\
         \midrule
         
          \multirow{10}{*}{{\large FT}} & \multirow{5}{*}{{\tt \large AB}} & SimCLR & StandardAug & $\mathbf{0.751}$ & $\mathbf{0.961}$ & $0.556$  & $\mathbf{0.883}$\\
          & & SimCLR & AugUS-O & $0.734$ & $0.960$ & $0.523$ &  $0.850$\\
          & & SimCLR & AugUS-D & $0.712$ & $0.934$ & $0.500$ &  $0.854$\\
          & & Random & - & $0.748$ & $0.885$ & $\mathbf{0.606}$ &  $0.853$\\
          & & ImageNet & - & $0.718$ & $0.903$ & $0.527$ &  $0.814$\\
          \cmidrule(l{3pt}){2-8} 
          & \multirow{5}{*}{{\tt \large PE}} & SimCLR & StandardAug & $\mathbf{0.805}$ & $0.838$ & $0.861$ &  $\mathbf{0.898}$\\
          & & SimCLR & AugUS-O & $0.572$ & $0.649$ & $0.717$ &  $0.536$\\
          & & SimCLR & AugUS-D & $0.800$ & $\mathbf{0.904}$ & $0.768$ &  $0.879$\\     
          & & Random & - & $0.700$ & $0.850$ & $0.643$ &  $0.804$\\
          & & ImageNet & - & $0.776$ & $0.775$ & $\mathbf{0.914}$ & $0.840$\\
         \bottomrule
    \end{tabular}
    }
    \label{tab:finetune-test-results-ext}
\end{table}

\subsection{Diagnostic Classification Task Evaluation}
\label{subsec:diagnostic-classification-task-evaluation}

Models pretrained on LUSData were also evaluated on the COVID-19 multi-class problem ({\tt COVID}).
Unlike the {\tt AB} and {\tt PE} tasks, {\tt COVID} is a diagnostic task that involves global image understanding, as the relationship between objects is pertinent.
Multiple findings on lung ultrasound have been observed in the context of COVID-19 pneumonia, including B-lines, pleural line abnormalities, and consolidation~\cite{blazic2023use}.

Linear classifiers were trained on the {\tt COVID} training set and evaluated on the {\tt COVID} test set. 
As shown in Table~\ref{tab:covid-test-set-results}, AugUS-O was observed to have the greatest test multiclass AUC, which was considered the primary metric of interest. 
Looking at the t-SNE visualizations in Fig.~\ref{fig:tsne}, AugUS-O corresponds to the only visualization where the representations for the COVID-19 Pneumonia and non-COVID-19 Pneumonia classes are clustered together.

Table~\ref{tab:covid-test-set-results} provides test metrics for fine-tuned {\tt COVID} classifiers.
Again, AugUS-O exhibited the best performance. 
Moreover, fine-tuned models generally performed worse than the linear classifiers trained on feature vectors from SSL-pretrained models; they likely suffered from overfitting, as COVIDx-US is a smaller dataset.

Unlike the {\tt AB} and {\tt PE} tasks, models trained for {\tt COVID} were pretrained using the LUSData dataset.
We repeated the linear classification and fine-tuning experiments using models pretrained on the COVIDx-US training set.
Table~\ref{tab:covid-test-set-results} reports the results of linear and fine-tuning evaluations.
As was observed for backbones pretrained on LUSData, pretraining with the AugUS-O pipeline resulted in the greatest test set AUC.

The trends observed for the {\tt COVID} task are different than those observed for the {\tt AB} and {\tt PE} tasks.
Regardless of the augmentation pipeline, SimCLR-pretrained weights resulted in better performance than ImageNet-pretrained or random weight initialization.
On object classification tasks, models pretrained using the StandardAug and AugUS-D pipelines performed the best.
However, on the diagnostic {\tt COVID} task, AugUS-O performed best.
Recall that AugUS-O was designed to retain semantic information, while both StandardAug and AugUS-D contain the very impactful crop \& resize (C\&R) transform that can obscure large portions of the image.
Object classification tasks require scale invariance, which is enforced by applying C\&R during SSL pretraining.
Diagnostic tasks, on the other hand, require global image context for interpreters to make a decision, which is preserved best by the AugUS-O pipeline.

\begin{table}[h!]
    \centering
    \caption{Test set performance for linear classification (LC) and fine-tuning (FT) experiments with the {\tt COVID} task. 
    Binary metrics are averages across classes.
    The best observed metrics in each experimental setting are in \textbf{boldface}.}
    \setlength{\tabcolsep}{0.125cm}
    {\small
    \begin{tabular}{ccccccccc}
        \toprule
         \makecell{Train \\ Setting} & \makecell{Pretraining \\ Dataset} & \makecell{Initial \\ Weights} & Pipeline & Accuracy & Precision & Recall & AUC \\
         \midrule
         \multirow{7}{*}{{\large LC}} & \multirow{3}{*}{{LUSData}} & SimCLR & StandardAug & $0.454$ & $0.371$ & $0.413$ & $0.784$ \\
          & & SimCLR & AugUS-O & $\mathbf{0.560}$ & $0.431$ & $\mathbf{0.513}$ &  $\mathbf{0.836}$ \\
          & & SimCLR & AugUS-D & $0.487$ & $0.348$ & $0.447$ & $0.713$ \\
        \cmidrule(l{3pt}){2-8} 
         & \multirow{3}{*}{{COVIDx-US}} & SimCLR & StandardAug & $0.498$ & $\mathbf{0.582}$ & $0.501$ & $0.781$ \\
          & & SimCLR& AugUS-O & $\mathbf{0.557}$ & $0.506$ & $\mathbf{0.543}$ & $\mathbf{0.820}$ \\
          & & SimCLR & AugUS-D & $0.540$ & $0.400$ & $0.491$ & $0.760$ \\
        \cmidrule(l{3pt}){2-8} 
        & - & ImageNet & - & $0.503$ & $0.304$ & $0.451$ & $0.699$\\
         \midrule

        \multirow{7}{*}{{\large FT}} & \multirow{3}{*}{LUSData} & SimCLR & StandardAug & $0.381$ & $0.259$ & $0.365$ & $0.753$ \\
         & & SimCLR & AugUS-O & $\mathbf{0.557}$ & $\mathbf{0.428}$ & $\mathbf{0.509}$ & $\mathbf{0.836}$\\
         & & SimCLR & AugUS-D & $0.465$ & $0.321$ & $0.430$ & $0.744$ \\
        \cmidrule(l{3pt}){2-8} 
         & \multirow{3}{*}{{COVIDx-US}} & SimCLR & StandardAug & $0.450$ & $\mathbf{0.540}$ & $0.464$ & $0.770$ \\
         & & SimCLR & AugUS-O & $0.517$ & $0.483$ & $\mathbf{0.510}$ & $\mathbf{0.814}$\\
         & & SimCLR & AugUS-D & $\mathbf{0.526}$ & $0.384$ & $0.479$ & $0.672$ \\
        \cmidrule(l{3pt}){2-8} 
         & - & Random & - & $0.423$ & $0.327$ & $0.401$ & $0.534$ \\
         & - & ImageNet & - & $0.502$ & $0.305$ & $0.457$ & $0.698$ \\
         \bottomrule
    \end{tabular}
    }
    \label{tab:covid-test-set-results}
\end{table}

\subsection{Object Detection Task Evaluation}

Recall that the {\tt PL} task is an object detection problem geared toward localizing the pleural line.
We evaluated the pretrained models on {\tt PL} to explore whether the trends observed for object-centric lung US classification tasks would hold for an object detection task, where locality understanding is explicit.
We considered two evaluation settings: one in which the pretrained backbones weights were held constant, and another in which the backbone's weights were trainable.
Table~\ref{tab:pl-obj-det} reports the average precision at a $50\%$ intersection over union threshold (AP@50) evaluated on the LUSData test set.
When the backbone weights were frozen, SimCLR pretraining with AugUS-O resulted in the greatest test AP@50.
These trends differed from the results observed for {\tt AB} and {\tt PE} classification, which both require object recognition.
We speculate that aggressive cropping during the pretraining phase likely produced positive pairs where one image contained a pleural line while the other did not, which we believe would make it difficult to learn representations for pleural line objects when pretraining with the StandardAug or AugUS-D pipelines.
The performance of the frozen pretrained backbones was strong overall, considering the low capacity of the backbone and the small, narrow shape of pleural line objects.
When fine-tuning end-to-end, SimCLR pretraining with AugUS-D resulted in the greatest test AP@50.

\begin{table}[h!]
    \centering
    \caption{LUSData local test set AP@50 for the {\tt PL} task observed for SSD models whose backbones were pretrained using different data augmentation pipelines.}
    \setlength{\tabcolsep}{0.3cm}
    \begin{tabular}{cccc}
        \toprule
         Backbone  & Initial Weights & Pipeline & AP@50 \\
         \midrule
          \multirow{5}{*}{\small Frozen} & SimCLR & StandardAug & $0.228$\\
          & SimCLR & AugUS-O & $\mathbf{0.255}$\\
          & SimCLR & AugUS-D &  $0.194$\\
          & Random & - &   $0.041$\\
          & ImageNet & - & $0.127$\\
        \midrule 
          \multirow{5}{*}{\small Trainable} & SimCLR & StandardAug & $0.316$\\
          & SimCLR & AugUS-O & $0.332$\\
          & SimCLR & AugUS-D &  $\mathbf{0.351}$\\
          & Random & - &   $0.308$\\
          & ImageNet & - & $0.310$\\
         \bottomrule
    \end{tabular}
    \label{tab:pl-obj-det}
\end{table}

\subsection{Label Efficiency Assessment}

\begin{figure}[h!]
    \centering
    \begin{subfigure}[b]{0.48\textwidth}
        \centering
        \includegraphics[height=8cm]{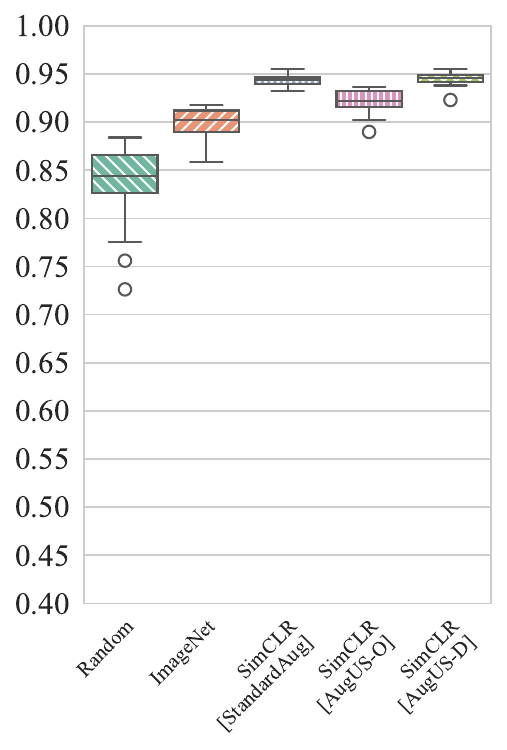}
        \caption{\textbf{(a)} {\tt AB} classification task}
        \label{subfig:label-eff-ab}
    \end{subfigure}
    \hfill
    \begin{subfigure}[b]{0.48\textwidth}
        \centering
        \includegraphics[height=8cm]{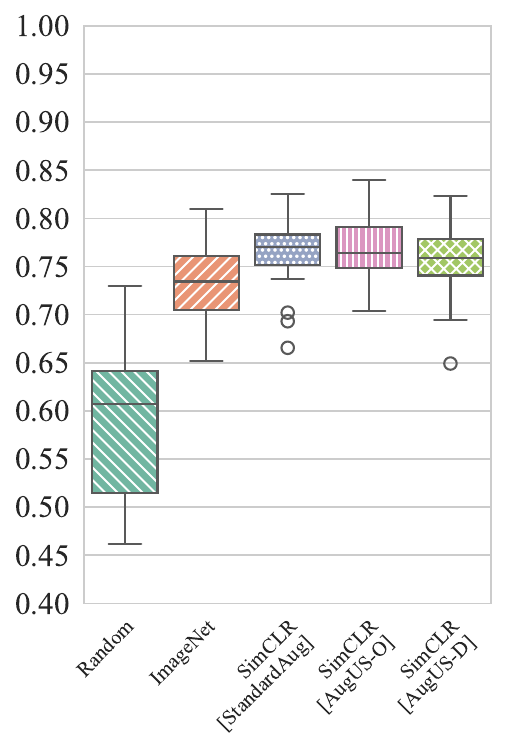}
        \caption{\textbf{(b)} {\tt PE} classification task}
        \label{subfig:label-eff-pe}
    \end{subfigure}
    \caption{Distribution of test AUC for classifiers trained on disjoint subsets of $5\%$ of the patients in the training partition of the private dataset}
    \label{fig:label-efficiency}
\end{figure}

Experiments were conducted to test the robustness of pretrained models in settings where few labeled samples are available.
The experiment was conducted only for the {\tt AB} and {\tt PE} classification tasks because there were enough unique videos and patients in the training set to create several disjoint subsets.
Backbones were fine-tuned on $20$ subsets of approximately $5\%$ of the LUSData training set, split by patient, yielding $20$ performance estimates for low-label settings.
Splitting was conducted at the patient level to heighten the difficulty of the task and to limit dependence between subsets.
Baseline estimates without SSL pretraining were obtained via initialization with random weights and with ImageNet-pretrained weights, resulting in five different performance conditions.
Figure~\ref{fig:label-efficiency} displays boxplots for test AUC distributions under each condition.
Friedman's Test indicated that there were significant differences among the median test AUC across conditions, for both the {\tt AB} and {\tt PE} tasks.
Post-hoc Wilcoxon Sign-Rank Tests were then conducted for each pair of conditions, using the Bonferroni correction with a family-wise error rate of $\alpha = 0.05$.
The median test AUCs of SimCLR-pretrained models were significantly greater than those initialized with random or ImageNet-pretrained weights for both the {\tt AB} and {\tt PE} tasks.
All medians were significantly different for {\tt AB}, except for the SimCLR-pretrained models using the StandardAug and AugUS-D pipelines, which achieved the greatest performance.
Notably, these pipelines both utilize the crop and resize transformation.
No significant differences were observed between any of the SimCLR-pretrained models for {\tt PE}.
\ref{apx:label-eff-stats} provides the test statistics for the above comparisons.

\subsection{Impact of Semantics-Preserving Preprocessing}

As outlined in Section~\ref{subsec:semantic-preserving-preprocessing}, all US images were cropped to the smallest rectangle enclosing the FOV because the areas outside the FOV are bereft of information.
Since pipelines containing the crop and resize transformation (C\&R) would be more likely to result in positive pairs that do not cover the FOV, it was hypothesized that cropping to the FOV as a preprocessing step would result in stronger pretrained backbones.
To investigate the effect of this semantics-preserving preprocessing step, we pretrained backbones on LUSData using each data augmentation pipeline and evaluated them on the {\tt AB}, {\tt AB}, and {\tt COVID} tasks.
Table~\ref{tab:semantic-preserving-preprocessing-ablation} compares the performance of each backbone with and without the preprocessing step.
Performance on the {\tt AB} task did not change.
However, test AUC on both the {\tt PE} and {\tt COVID} tasks was consistently lower when the semantics-preserving preprocessing was not applied.
Note that greatly less labelled data is available for {\tt PE} and {\tt COVID} than for {\tt AB}.
Based on these experiments, FOV cropping is a valuable semantics-preserving preprocessing step for multiple LUS classification tasks.

\begin{table}[]
    \caption{Test set AUC for SimCLR-pretrained models with (\cmark) and without (\xmark) semantics-preserving preprocessing. 
    Results are reported for linear classifiers and fine-tuned models.}
    \label{tab:semantic-preserving-preprocessing-ablation}
    \centering
    \begin{tabular}{cccccc}
        \toprule
        & & \multicolumn{2}{c}{Linear classifier} & \multicolumn{2}{c}{Fine-tuned} \\
        \cmidrule(lr){3-4} \cmidrule(lr){5-6}
         Task & Pipeline / Preprocessing & \xmark & \cmark & \xmark & \cmark  \\
         \midrule
          \multirow{3}{*}{{\tt AB}} & StandardAug & $0.971$ & $0.970$ & $0.971$ & $0.970$ \\
          & AugUS-O & $0.950$ & $0.953$ & $0.926$ & $0.968$ \\
          & AugUS-D & $0.971$ & $0.971$ & $0.961$ & $0.962$ \\
        \midrule 
          \multirow{3}{*}{{\tt PE}} &  StandardAug & $0.873$ & $0.893$ & $0.869$ & $0.882$ \\
          &  AugUS-O & $0.846$ & $0.865$ & $0.522$ & $0.557$ \\
          &  AugUS-D & $0.867$ & $0.897$ & $0.864$ & $0.884$ \\
         \midrule 
          \multirow{3}{*}{{\tt COVID}}  & StandardAug & $0.742$ & $0.784$ & $0.724$ & $0.753$ \\
           & AugUS-O & $0.793$ & $0.836$ & $0.805$ & $0.836$ \\
           & AugUS-D & $0.585$ & $0.713$ & $0.737$ & $0.744$ \\
         \bottomrule
    \end{tabular}
\end{table}

\subsection{Impact of the Cropping in Object Classification Tasks}

\begin{figure}[h!]
    \centering
    \includegraphics[width=0.7\linewidth]{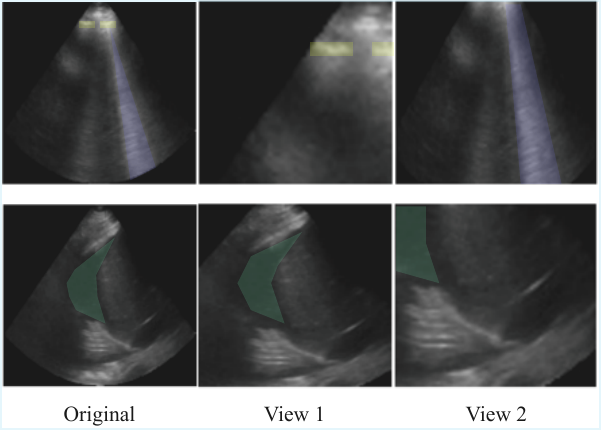}
    \caption{Examples of how the random crop and resize transformation (B00) can reduce semantic information. 
    Original images are on the left, and two random crops of the image are on the right. 
    \textbf{Top:} The original image contains a B-line (purple), which is visible in View 2 but not in View 1. 
    The original image also contains instances of the pleural line (yellow) which are visible in View 1 but not in View 2.
    \textbf{Bottom:} The original image contains a pleural effusion (green), which is visible in View 1 but largely obscured in View 2. }
    \label{fig:crop_examples}
\end{figure}

The leave-one-out analysis for transformations exhibited the striking finding that crop and resize (C\&R) was the most effective transformation in the StandardAug pipeline for the two object classification tasks: {\tt AB} and {\tt PE}. 
Moreover, both pipelines containing C\&R resulted in the greatest downstream test performance on {\tt AB} and {\tt PE}.
Ordinarily, crops are taken at random locations in an image, with areas between $8\%$ and $100\%$ of the original image's area.
Aggressive crops can create situations in which positive pairs do not contain the same objects of interest.
Fig.~\ref{fig:crop_examples} shows how C\&R could produce positive pairs with different semantic content.
Despite this, the results indicated that pipelines containing C\&R led to improved performance for the object-centric {\tt AB} and {\tt PE} tasks.
The exceptional influence of C\&R warranted further investigation into its hyperparameters.

\begin{figure}[h!]
    \centering
    \begin{subfigure}[b]{0.48\textwidth}
        \centering
        \includegraphics[height=5cm]{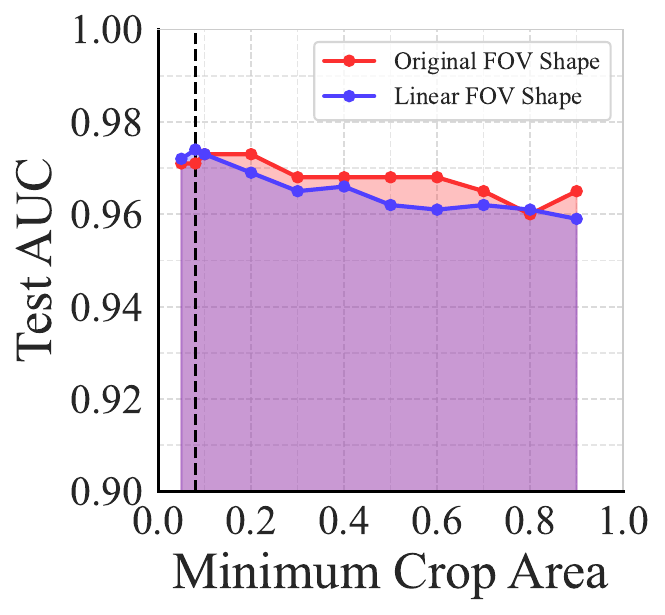}
        \caption{\textbf{(a)} {\tt AB} classification task}
        \label{subfig:crop-ablation-ab}
    \end{subfigure}
    \hfill
    \begin{subfigure}[b]{0.48\textwidth}
        \centering
        \includegraphics[height=5cm]{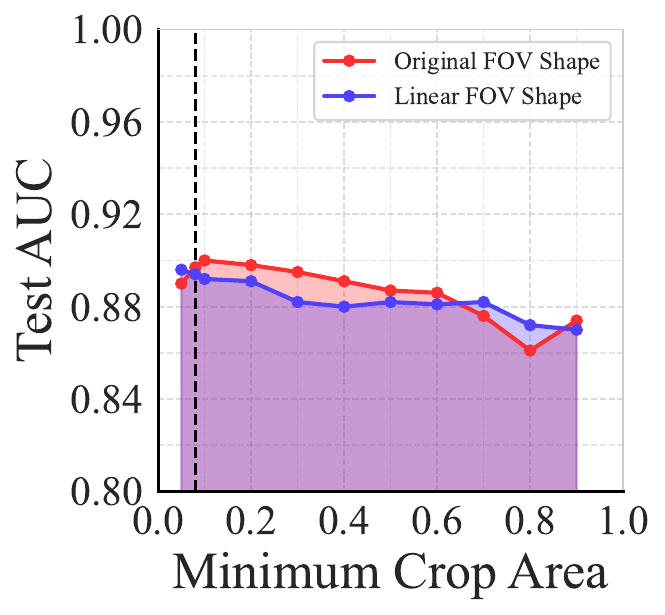}
        \caption{\textbf{(b)} {\tt PE} classification task}
        \label{subfig:crop-ablation-pe}
    \end{subfigure}
    \caption{Test set AUC for linear classifiers trained on the representations outputted by pretrained backbones.
    Each backbone was pretrained using AugUS-D with different values for the minimum crop area, $c$. 
    Results are provided for models pretrained with the original ultrasound FOV, along with images transformed to linear FOV only. 
    The dashed line indicates the default value of $c=0.08$.}
    \label{fig:crop_results}
\end{figure}

We investigated the impact of the minimum crop area, $c$, as a hyperparameter.
Models were pretrained with the AugUS-D pipeline, using values for $c$ in the range $[0.05, 0.9]$.
Linear evaluation was conducted for the {\tt AB} and {\tt PE} tasks.
As shown in Fig.~\ref{fig:crop_results}, smaller values of $c$ yielded better performance, peaking at $c \approx 0.1$.
The default assignment of $c=0.08$ was already a reasonable choice for these two tasks.
Additional experiments elucidating the effects of C\&R hyperparameters can be found in \ref{apx:rcr-experiments}.

Another concern with C\&R is that it could result in crops covering the black background on images with a convex FOV.
Despite the semantics-preserving preprocessing (described in Section~\ref{fig:semantic-preprocessing}), the top left and right corners of such images provide no information.
To characterize the robustness of pretraining under these circumstances, we repeated the experiments sweeping over $c \in [0.05, 0.9]$ but first applied the probe type change transformation (i.e., U00) to every convex FOV.
Thus, all inputs to the model were linear FOVs devoid of non-semantic background.
A by-product of this transformation is that the near fields of convex images are horizontally stretched.
As seen in Fig.~\ref{fig:crop_results}, this change resulted in a slight decrease in performance for both the {\tt AB} and {\tt PE} tasks.
Evidently, the detriment of spatial distortion outweighed the benefit of guaranteeing that crops were positioned over semantic regions.

Overall, it is clear that aggressive C\&R is beneficial for distinguishing between A-lines and B-lines and detecting pleural effusions on lung US.
Both are object-centric classification tasks.
Even though some crops may not contain the object, the backbone would be exposed to several paired instances of transformed portions of objects during pretraining, potentially facilitating texture and shape recognition.
Conversely, solving diagnostic tasks such as {\tt COVID} requires a holistic assessment of the FOV, wherein the co-occurrence of objects is contributory to the overall impression.

\section{Conclusion}
\label{sec:conclusion}

This study proposed data augmentation and preprocessing strategies for self-supervised learning in ultrasound.
A commonly employed baseline pipeline (StandardAug) was compared to a handcrafted semantics-preserving pipeline (AugUS-O) and a hybrid pipeline (AugUS-D) composed from the first two.
Evaluation on lung US interpretation tasks revealed a dichotomy between the utility of the pipelines.
Pipelines featuring the cropping transformation (StandardAug and AugUS-D) were most useful for object classification and detection tasks in LUS.
On the other hand, AugUS-O -- designed to preserve semantics in LUS -- resulted in the greatest performance on a diagnostic task that required global context.
Additionally, US field of view cropping was found to be a beneficial preprocessing step for multiple lung US classification tasks, regardless of the data augmentation strategy.
Based on the results, developers should use semantics-preserving preprocessing during pretraining.
Regarding data augmentation, semantics-preserving transformations should be considered for tasks requiring holistic interpretation of images, cropping-based transformations should be leveraged for object-centric downstream tasks.

Some limitations are acknowledged in this study.
For example, SimCLR was the only SSL objective that was investigated, and all downstream tasks were confined to the lung.
Moreover, some of the transformations introduced in this work constitute computationally expensive preprocessing steps, as they are applied with nonzero probability to each image.

Future work should apply this study's methods to assess the impact of data augmentation pipelines for US diagnostic tasks outside of the lung and for other SSL methods.
Future studies could also compare data augmentation strategies for localization and segmentation downstream tasks in ultrasound.

\section*{CRediT authorship contribution statement}

\textbf{Blake VanBerlo:} Conceptualization, Methodology, Software, Validation, Formal Analysis, Investigation, Data Curation, Writing - Original Draft, Writing - Review \& Editing, Visualization, Project Administration, Funding Acquisition
\textbf{Jesse Hoey:} Conceptualization, Writing - Review \& Editing, Supervision
\textbf{Alexander Wong:} Conceptualization, Writing - Review \& Editing, Supervision
\textbf{Robert Arntfield:} Resources, Data Curation, Writing - Review \& Editing

\section*{Declaration of competing interest}

There are no competing interests for all authors.

\section*{Data availability}

The LUSData dataset is private.
COVIDxUS~\cite{Ebadi2022-mn} is publicly available.
Code will be published in a public GitHub repository upon publication.

\section*{Acknowledgements}

This work was supported by the Natural Sciences and Engineering Research Council of Canada, as Blake VanBerlo is a Vanier Scholar (FRN 186945). 
Computational resource support was also provided by Compute Ontario (computeontario.ca) and the Digital Research Alliance of Canada \\ (alliance.can.ca).

\appendix

\section{Dataset Details}
\label{apx:dataset-details}

This section provides further details regarding the composition of the LUSData and COVIDx-US datasets, stratified by different attributes. 
Table~\ref{tab:lusdata-characteristics} provides a breakdown of the characteristics of the unlabelled data, training set, validation set, local test set, and external test set.
As can be seen in the table, the external test set's manufacturer and probe type distributions differ greatly from those in the local test set.
As such, the external test set constitutes a meaningful assessment of generalizability for models trained using the LUSData dataset.
Table~\ref{tab:covidxus-characteristics} provides similar information for the COVIDx-US dataset~\cite{Ebadi2022-mn}.
Note that the probe types listed in Tables~\ref{tab:lusdata-characteristics}~and~\ref{tab:covidxus-characteristics} are predictions produced by a product and not from meta-data accompanying the examinations.\footnote{Product name withheld to preserve anonymity during peer review}.

\begin{table*}[h!]
    \centering
    \caption{Characteristics of the ultrasound videos in the LUSData dataset. 
    The number of videos possessing known values for a variety of attributes are displayed.
    Percentages of the total in each split are provided as well, but some do not sum to $100$ due to rounding.}
    {\scriptsize
    \begin{threeparttable}
    \begin{tabular}{p{2.5cm}lllll}
        \toprule
         & \multicolumn{4}{c}{Local} & External \\
         \cmidrule(r){2-5} \cmidrule(l){6-6}
         & Unlabeled & Train & Validation & Test & Test \\
         \midrule
         \multicolumn{1}{l}{\textbf{Probe Type}} &  &  &  &  &  \\ 
         \multicolumn{1}{r}{Phased Array} & \num{50769}$\,(85.6\%)$ & $5146\,(90.6\%)$ & $1051\,(88.8\%)$ & $1062\,(85.0\%)$ & $586\,(63.4\%)$ \\ 
         \multicolumn{1}{r}{Curved Linear} & $6601\,(11.1\%)$ & $439\,(7.7\%)$ & $108\,(9.1\%)$ & $167\,(13.4\%)$ & $92\,(9.9\%)$ \\ 
         \multicolumn{1}{r}{Linear} & $1939\,(3.3\%)$ & $94\,(1.7\%)$ & $25\,(2.1\%)$ & $20\,(1.6\%)$ & $247\,(26.7\%)$ \\[8pt] 
         
         \multicolumn{1}{l}{\textbf{Manufacturer}} &  &  &  &  &  \\ 
         \multicolumn{1}{r}{Sonosite} & \num{53663}$\,(90.5\%)$ & $4386\,(77.2\%)$ & $848\,(71.6\%)$ & $963\,(77.1\%)$ & $626\,(67.7\%)$ \\ 
         \multicolumn{1}{r}{Mindray} & $4045\,(6.8\%)$ & $847\,(14.9\%)$ & $216\,(18.2\%)$ & $153\,(12.2\%)$ & $55\,(5.9\%)$ \\ 
         \multicolumn{1}{r}{Philips} & $66\,(0.1\%)$ & $50\,(0.9\%)$ & $6\,(0.5\%)$ & $11\,(0.9\%)$ & $244\,(26.4\%)$ \\ 
         \multicolumn{1}{r}{Esaote} & $233\,(0.4\%)$ & $4\,(0.1\%)$ & $0\,(0.0\%)$ & $0\,(0.0\%)$ & $0\,(0.0\%)$ \\ 
         \multicolumn{1}{r}{GE\tnote{\S}} & $10\,(0.0\%)$ & $0\,(0.0\%)$ & $0\,(0.0\%)$ & $0\,(0.0\%)$ & $0\,(0.0\%)$ \\[8pt] 
         
         \multicolumn{1}{l}{\textbf{Depth (cm)}} &  &  &  &  &  \\ 
         \multicolumn{1}{r}{Mean [STD]} & $14.3\,[4.5]$ & $13.0\,[3.8]$ & $13.1\,[3.7]$ & $12.7\,[3.8]$ & $11.1\,[4.8]$ \\ 
         \multicolumn{1}{r}{Unknown} & $1606\,(2.7\%)$ & $407\,(7.2\%)$ & $115\,(9.7\%)$ & $122\,(9.8\%)$ & $0\,(0.0\%)$ \\[8pt]  
         
         \multicolumn{1}{l}{\textbf{Environment}} &  &  &  &  &  \\ 
         \multicolumn{1}{r}{ICU\tnote{\textdagger}} & \num{43839}$\,(73.9\%)$ & $3722\,(65.5\%)$ & $706\,(59.6\%)$ & $727\,(58.2\%)$ & $0\,(0.0\%)$ \\ 
         \multicolumn{1}{r}{ER\tnote{\textdaggerdbl}} & \num{13280}$\,(22.4\%)$ & $760\,(13.4\%)$ & $206\,(17.4\%)$ & $253\,(20.3\%)$ & $0\,(0.0\%)$ \\ 
         \multicolumn{1}{r}{Ward} & $2033\,(3.4\%)$ & $173\,(3.0\%)$ & $25\,(2.1\%)$ & $49\,(3.9\%)$ & $0\,(0.0\%)$ \\ 
         \multicolumn{1}{r}{Urgent Care} & $129\,(0.2\%)$ & $3\,(0.1\%)$ & $0\,(0.0\%)$ & $1\,(0.1\%)$ & $0\,(0.0\%)$ \\ 
         \multicolumn{1}{r}{Unknown} & $28\,(0.0\%)$ & $1021\,(18.0\%)$ & $247\,(20.9\%)$ & $219\,(17.5\%)$ & $925\,(100.0\%)$ \\[8pt] 
         
         \multicolumn{1}{l}{\textbf{Patient Sex}} &  &  &  &  &  \\ 
         \multicolumn{1}{r}{Male} & \num{30300}$\,(51.1\%)$ & $2963\,(52.2\%)$ & $607\,(51.3\%)$ & $588\,(47.1\%)$ & $0\,(0.0\%)$ \\ 
         \multicolumn{1}{r}{Female} & \num{20809}$\,(35.1\%)$ & $1793\,(31.6\%)$ & $325\,(27.4\%)$ & $412\,(33.0\%)$ & $0\,(0.0\%)$ \\ 
         \multicolumn{1}{r}{Unknown} & $8200\,(13.8\%)$ & $923\,(16.3\%)$ & $252\,(21.3\%)$ & $249\,(19.9\%)$ & $925\,(100.0\%)$ \\[8pt]  
         
         \multicolumn{1}{l}{\textbf{Patient Age}} &  &  &  &  &  \\ 
         \multicolumn{1}{r}{Mean [STD]} & $62.3\,[20.0]$ & $63.3\,[16.5]$ & $62.0\,[18.4]$ & $62.8\,[17.3]$ & $-$ \\ 
         \multicolumn{1}{r}{Unknown} & $53\,(0.1\%)$ & $1029\,(18.2\%)$ & $247\,(20.9\%)$ & $219\,(17.5\%)$ & $925\,(100.0\%)$ \\ 
         \midrule
         \multicolumn{1}{l}{\textbf{Total}} & $\num{59309}$ & $5679$ & $1184$ & $1249$ & $925$ \\ 
         \bottomrule
    \end{tabular}
    \begin{tablenotes}
    \footnotesize
    \item[\S] General Electric
    \item[\textdagger] Intensive Care Unit
    \item[\textdaggerdbl] Emergency Room
    \end{tablenotes}
    \end{threeparttable}}
    \label{tab:lusdata-characteristics}
\end{table*}

\begin{table*}[h!]
    \centering
    \caption{Characteristics of the ultrasound videos in the COVIDx-US dataset. 
    The number of videos possessing known values for a variety of attributes are displayed.
    Percentages of the total in each split are provided as well, but some do not sum to $100$ due to rounding.}
    {\footnotesize
    \setlength{\tabcolsep}{0.4cm}
    \begin{tabular}{p{4cm}lll}
        \toprule
         & Train & Validation & Test  \\
         \midrule
         \multicolumn{1}{l}{\textbf{Probe Type}}   &  &  &    \\ 
         \multicolumn{1}{r}{Phased Array} &  $55\,(32.5\%)$ & $18\,(42.9\%)$ & $11\,(35.5\%)$ \\ 
         \multicolumn{1}{r}{Curved Linear}  & $83\,(49.1\%)$ & $13\,(31.0\%)$ & $10\,(32.3\%)$  \\ 
         \multicolumn{1}{r}{Linear}  & $31\,(18.3\%)$ & $11\,(26.2\%)$ & $10\,(32.3\%)$  \\[8pt] 
         
         \multicolumn{1}{l}{\textbf{Patient Sex}} &  &  &    \\ 
         \multicolumn{1}{r}{Male}  & $33\,(19.5\%)$ & $15\,(35.7\%)$ & $7\,(22.6\%)$ \\ 
         \multicolumn{1}{r}{Female}  & $18\,(10.7\%)$ & $3\,(7.1\%)$ & $3\,(9.7\%)$  \\ 
         \multicolumn{1}{r}{Unknown}  & $118\,(69.8\%)$ & $24\,(57.1\%)$ & $21\,(67.7\%)$ \\[8pt]  
         
         \multicolumn{1}{l}{\textbf{Patient Age}} &    &  &    \\ 
         \multicolumn{1}{r}{Mean [STD]} & $36.6\,[14.6]$ & $47.6\,[18.9]$ & $52.9\,[21.1]$ \\ 
         \multicolumn{1}{r}{Unknown}  & $127\,(75.1\%)$ & $22\,(52.4\%)$ & $20\,(64.5\%)$ \\ 
         \midrule
         \multicolumn{1}{l}{\textbf{Total}}   & $169$ & $42$  & $31$ \\ 
         \bottomrule
    \end{tabular}}
    \label{tab:covidxus-characteristics}
\end{table*}

\section{Transformation Runtime Estimates}
\label{apx:transformation-runtime-estimates}

We aimed to examine relative runtime differences between the transformations used in this study.
Runtime estimates were obtained for each transformation in the StandardAug and AugUS-v1 pipelines.
Estimates were calculated by conducting the transformation $1000$ times using the same image.
The experiments were run on a system with an Intel i9-10900K CPU at \SI{3.7}{GHz}.
Python 3.11 was utilized, and the transforms were written using PyTorch version 2.2.1 and TorchVision 0.17.1.
Note that runtime may vary considerably depending on the software environment and underlying hardware.

\section{StandardAug Transformations}
\label{apx:byol-transformations}

We investigated a standard data augmentation pipeline that has used extensively in the SSL literature~\cite{chen2020simple,grill2020bootstrap,zbontar2021barlow,bardes2022vicreg}.
To standardize experiments, we adopted the symmetric version from the VICReg paper~\cite{bardes2022vicreg}, which uses the same transformations and probabilities for each of the two branches in the joint embedding architecture. 
The transformations and their parameter settings are widely adopted.
Below, we detail their operation and parameter assignments.

\subsection{Crop and Resize (B00)}
A rectangular crop of the input image is designated at a random location.
The area of the cropped region is sampled from the uniform distribution $\mathcal{U}(0.08, 1)$.
The cropped region's aspect ratio is sampled from the uniform distribution $\mathcal{U}(0.75, 1.33)$.
Its width and height are calculated accordingly.
The cropped region is then resized to the original image dimension.

\subsection{Horizontal Reflection (B01)}

The image is reflected about the central vertical axis.

\subsection{Color Jitter (B02)}

The brightness, contrast, saturation, and hue of the image are modified.
The brightness change factor, contrast change factor, saturation change factor, and hue change factor are sampled from $\mathcal{U}(0.6, 1.4)$, $\mathcal{U}(0.6, 1.4)$, $\mathcal{U}(0.8, 1.2)$, and $\mathcal{U}(-0.1, 0.1)$, respectively.

\subsection{Conversion to Grayscale (B03)}

Images are converted to grayscale.
The output images have three channels, such that each channel has the same pixel intensity.

\subsection{Gaussian Blur (B04)}

The image is denoised using a Gaussian blur with kernel size $13$ and standard deviation sampled uniformly at random from $\mathcal{U}(0.1, 2)$.
Note that in the original pipeline, the kernel size was set to $23$ and $224 \times 224$ images were used. 
We used $128 \times 128$ images; as such, we selected a kernel size that covers a similar fraction of the image.

\subsection{Solarization (B05)}

All pixels with intensity of $128$ or greater are inverted.
Note that the inputs are unsigned 8-bit images.

\section{Ultrasound-Specific Transformations}
\label{apx:us-specific-transformations}

In this section, we provide details on the set of transformations that comprise AugUS-v1.

\begin{figure}[h!]
    \centerline{\includegraphics[width=0.7\columnwidth]{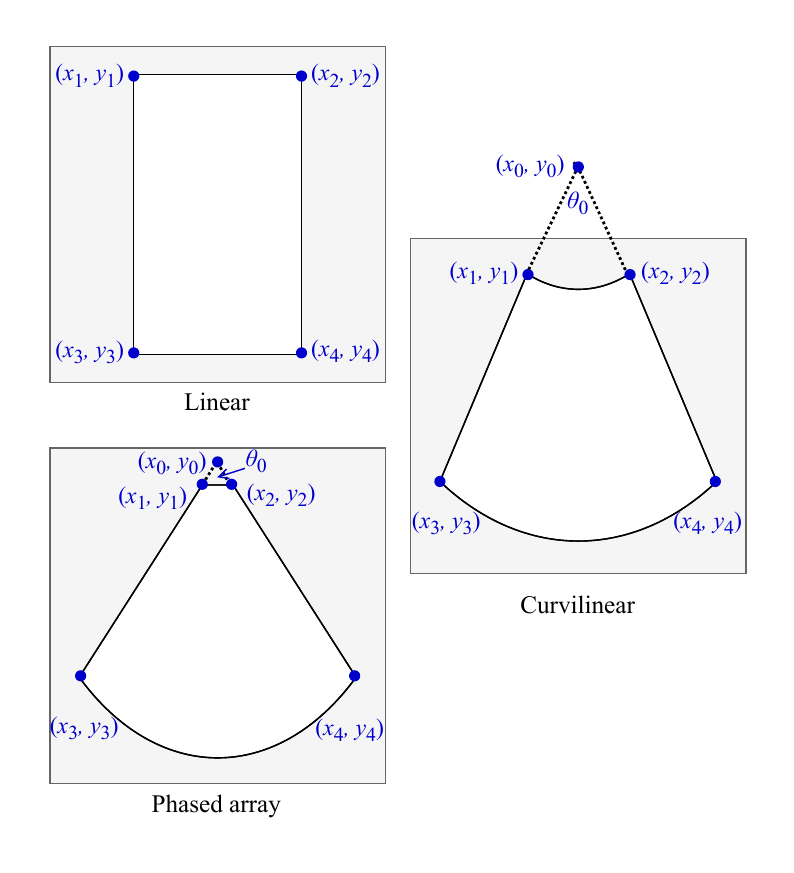}}
    \caption{Locations of the named FOV vertices for each of the three main field of view shapes in US imaging}
    \label{fig:beam-vertices}
\end{figure}

Several of the transformations operate on the pixels contained within the ultrasound field of view (FOV).
As such, the geometrical form of the FOV was required to perform some transformations.
We adopted the same naming convention for the vertices of the ultrasound FOV as Kim~\etal~\cite{kim2023point}.
Let $p_1$, $p_2$, $p_3$, and $p_4$ represent the respective locations of the top left, top right, bottom left, and bottom right vertices, and let $\langle x_i, y_i \rangle$ be the x- and y-coordinates of $p_i$ in image space.
For convex FOV shapes, we denote the intersection of lines $\overleftrightarrow{p_1p_3}$ and $\overleftrightarrow{p_2p_4}$ as $p_0$.
Fig.~\ref{fig:beam-vertices} depicts the arrangement of these points for each of the three main ultrasound FOV shapes: linear, curvilinear, and phased array.
A software tool was used to estimate the FOV shape and probe type for all videos in each dataset (UltraMask, Deep Breathe Inc., Canada).

\subsection{Probe Type Change (U00)}

To produce a transformed ultrasound image with a different FOV shape, a mapping that gives the location of pixels in the original image for each coordinate in the new image is calculated.
Concretely, the function $\mathbf{f}: \mathbb{R}^2 \rightarrow [-1, 1]^2$ returns the coordinates of the point in the original image that corresponds to a point in the transformed image.
Note that $(-1,-1)$ corresponds to the top left of the original image.
Pixel intensities in the transformed image are interpolated according to their corresponding location in the original image.

Algorithm~\ref{alg:linear-to-convex} details the calculation of $\mathbf{f}_{\ell \rightarrow c}$ for converting linear FOVs to convex FOVs with a random radius factor $\rho \sim \mathcal{U}(1, 2)$, along with new FOV vertices.
Similarly, curvilinear and phased array FOV shapes are converted to linear FOV shapes.
Algorithm~\ref{alg:convex-to-linear} details the calculation of the mapping $\mathbf{f}_{c \rightarrow \ell}$ that transforms convex FOV shapes to linear shapes, along with calculations for the updated named coordinates.
To ensure that no aspects of the old FOV remain on the image, bitmask $M' \in \{0, 1\}^{h \times w}$ is produced using the new named coordinates.

Since the private dataset was resized to square images that exactly encapsulated the FOV, images are resized to match their original aspect ratios to ensure that the sectors were circular.
They are then resized to their original dimensions following the transformation.

\begin{algorithm}
\begin{spacing}{1.2} 
\caption{Compute a point mapping for linear to curvilinear FOV shape, along with new FOV vertices}
\label{alg:linear-to-convex}
\begin{algorithmic}[1]
\REQUIRE FOV vertices $p_1$, $p_2$, $p_3$, $p_4$; radius factor $\rho$; coordinate maps $\mathbf{x} = \mathbf{1}_{h \times 1} [0, 1, \ldots, w-1]$ and $\mathbf{y} = [0, 1, \ldots, h-1]^T \mathbf{1}_{1 \times w} $
\STATE $r_b  \gets \rho (y_3 - y_1)$ \CustomComment{Bottom sector radius}
\STATE $x_0'  \gets \max(x_3, 0) + (x_4 - x_3) / 2$ 
\STATE $y_0'  \gets y_3 - r_b$  \CustomComment{Lateral bounds will intersect at $(x_0',y_0')$}
\STATE $y_1' = y_2'  \gets y_1$
\STATE $y_3' = y_4'  \gets y_0' +  \sqrt{r_b^2 - (x_0' - x_1)^2}$
\STATE $x_1'  \gets x_0' - (y_1 - y_0') (x_0' - x_3) / (y_3' - y_0')$ 
\STATE $x_2'  \gets x_0' + (y_1 - y_0') (x_0' - x_3) / (y_3' - y_0')$ 
\STATE $r_t  \gets \sqrt{(x_0' - x_1')^2 + (y_1 - y_0')^2}$ \CustomComment{Top sector radius}
\STATE $\bm{\phi} \gets \atantwo (\mathbf{x} - x_0', \mathbf{y} - y_0')$ \CustomComment{Angle with the central vertical}
\renewcommand{\arraystretch}{1.5}
\STATE $\mathbf{f}_{\ell \rightarrow c} \gets \begin{pmatrix}
     \frac{\bm{\phi} + (x_i - w/2)/w}{|\,\bm{\phi}[y_3',0]\,|}  \\[7pt]
     2 \frac{\sqrt{(x_0' - \mathbf{x})^2 + (y_0' - \mathbf{y})^2} - r_t}{r_b - r_t} - 1 \end{pmatrix}$ \CustomComment{Final coordinate mapping}
\renewcommand{\arraystretch}{1.0}
\RETURN $\mathbf{f}_{\ell \rightarrow c}$, $p_1'$, $p_2'$, $p_3'$, $p_4'$ \CustomComment{Coordinate mapping, new FOV vertices}
\end{algorithmic}
\end{spacing}
\end{algorithm}

\begin{algorithm}
\begin{spacing}{1.2} 
\caption{Compute a point mapping for convex to linear FOV shape, along with new FOV vertices}
\label{alg:convex-to-linear}
\begin{algorithmic}[1]
\REQUIRE FOV vertices $p_1$, $p_2$, $p_3$, $p_4$; point of intersection $p_0$; angle of intersection $\theta_0$; width fraction $\omega \in [0, 1]$;  coordinate maps \\ $\mathbf{x} = \mathbf{1}_{h \times 1} [0, 1, \ldots, w-1]$ and $\mathbf{y} = [0, 1, \ldots, h-1]^T\mathbf{1}_{1 \times w} $
\STATE $r_b  \gets \sqrt{(x_0 - x_3)^2 + (y_0 - y_3)^2}$ \CustomComment{Bottom sector radius}
\STATE $x_1' = x_3' \gets x_0 - \omega w / 2$ 
\STATE $x_2' = x_4' \gets x_0 + \omega w / 2$ 
\STATE $y_1' = y_2' \gets y_1$ 
\STATE $y_3' = y_4' \gets y_0 + r_b$ 

\STATE $\boldsymbol{\phi} \gets \theta_0 ((\mathbf{x} - x_3) / (x_4 - x_3) - \frac{1}{2})$ \CustomComment{Angle with the central vertical}
\STATE $\mathbf{y_n} \gets (\mathbf{y} - y_1) / (y_0 + r_b - y_1)$ \CustomComment{Normalized y-coordinates}

\IF{probe type is curvilinear} 
    \STATE $r_t \gets \sqrt{(x_0 - x_1)^2 + (y_0 - y_1)^2}$ \CustomComment{Curvilinear; top sector radius}
\ELSE
    \STATE $r_t \gets (y_1 - y_0) / \cos{(\boldsymbol{\phi} / w)}$ \CustomComment{Phased array; distance to top bound}
    \renewcommand{\arraystretch}{1.2}
\ENDIF
\renewcommand{\arraystretch}{1.2}
\STATE $\mathbf{f}_{c \rightarrow \ell} \gets \begin{pmatrix}
 x_0 + \sin{(\boldsymbol{\phi} / w)} (r_t + \mathbf{y_n} (r_b - r_t))  \\
 y_0 + \cos{(\boldsymbol{\phi} / w)} (r_t + \mathbf{y_n} (r_b - r_t))  \end{pmatrix}$
\renewcommand{\arraystretch}{1.0}

\RETURN $\mathbf{f}_{c \rightarrow \ell}$, $p_1'$, $p_2'$, $p_3'$, $p_4'$ \CustomComment{Coordinate mapping, new FOV vertices}
\end{algorithmic}
\end{spacing}
\end{algorithm}

\subsection{Convexity Change (U01)}

To mimic an alternative convex FOV shape with a different $\theta_0$, a mapping is calculated that results in a new FOV shape wherein $p_0$ is translated vertically.
A new value for the width of the top of the FOV is randomly calculated, facilitating the specification of a new $p_0$.
Given the new $p_0$, a pixel map $\mathbf{f}_{c \rightarrow c'}$ is computed according to Algorithm~\ref{alg:convexity-change}.
Similar to the probe type change transformation, pixel intensities at each coordinate in the transformed image are interpolated according to the corresponding coordinate in the original image returned by $\mathbf{f}_{c \rightarrow c'}$.

\begin{algorithm}
\begin{spacing}{1.2} 
\caption{Compute a point mapping from an original to a modified convex FOV shape.}
\label{alg:convexity-change}
\begin{algorithmic}[1]
\REQUIRE FOV vertices $p_1$, $p_2$, $p_3$, $p_4$; point of intersection $p_0$; angle of intersection $\theta_0$; new top width $w'$; coordinate maps \\ $\mathbf{x} = \mathbf{1}_{h \times 1} [0, 1, \ldots, w-1]$ and  $\mathbf{y} = [0, 1, \ldots, h-1]^T\mathbf{1}_{1 \times w} $
\STATE $s \gets w' (x_4 - x_3) / (x_2 - x_1)$ \CustomComment{Scale change for top bound}
\STATE $x_1' \gets x_0 - s(x_0 - x_1)$ 
\STATE $x_2' \gets x_0 + s(x_2 - x_0)$ 
\STATE $y_1' = y_2' \gets y_1$ 
\STATE $y_3' = y_4' \gets y_3$ 
\STATE $\theta_0', p_0' \gets$ angle, point of intersection of $\overleftrightarrow{p_1'p_3'}$ and $ \overleftrightarrow{p_2'p_4'}$

\STATE $r_b \gets \sqrt{(x_0 - x_3)^2 + (y_0 - y_3)^2}$ \CustomComment{Current bottom radius}
\STATE $r_b' \gets \sqrt{(x_0' - x_3')^2 + (y_0' - y_3')^2}$ \CustomComment{New bottom radius}
\STATE $r_t \gets \sqrt{(x_0 - x_1)^2 + (y_0 - y_1)^2}$ \CustomComment{Current top radius}
\STATE $r_t' \gets \sqrt{(x_0' - x_1')^2 + (y_0' - y_1')^2}$ \CustomComment{New top radius}

\STATE $\boldsymbol{\phi}' \gets \atantwo (\mathbf{x} - x_0', \mathbf{y} - y_0')$
\STATE $\mathbf{r}' \gets \sqrt{(x_0' - \mathbf{x})^2 + (y_0' - \mathbf{y})^2}$ 
\STATE $\mathbf{r}' \gets (\mathbf{r}' - r_t')(r_b - r_t)/(r_b' - r_t') + r_t$

\renewcommand{\arraystretch}{1.2}
\STATE $\mathbf{f}_{c \rightarrow c'}(x, y) \gets \begin{pmatrix}
 x_0 + \mathbf{r}' \sin{(\boldsymbol{\phi}' \theta_0 / \theta_0')}  \\
 y_0 + \mathbf{r}' \cos{(\boldsymbol{\phi}' \theta_0 / \theta_0')}   \end{pmatrix}$
\renewcommand{\arraystretch}{1.0}

\RETURN $\mathbf{f}_{c \rightarrow c'}$, $p_1'$, $p_2'$, $p_3'$, $p_4'$ \CustomComment{Coordinate mapping, new FOV vertices}
\end{algorithmic}
\end{spacing}
\end{algorithm}

\subsection{Wavelet Denoising (U02)}

Following the soft thresholding method by Birg{\'e} and Massart~\cite{birge1997model}, we apply a wavelet transform, conduct thresholding, then apply the inverse wavelet transform.
The mother wavelet is randomly chosen from a set, and the sparsity parameter $\alpha$ is sampled from a uniform distribution.
We use $J=3$ levels of wavelet decomposition and set the decomposition level $J_0=2$.
We designated Daubechies wavelets $\{ \mathit{db2}, \mathit{db5} \}$ as the set of mother wavelets, which is a subset of those identified by Vilimek~\etal's assessment~\cite{vilimek2022comparative} as most suitable for denoising US images.

\subsection{Contrast-Limited Adaptive Histogram Equalization (U03)}

Contrast-limited adaptive histogram equalization is applied to the input image.
The transformation enhances low-contrast regions of ultrasound images while avoiding excessive noise amplification.
We found that CLAHE enhances artifact 
The tiles are $8 \times 8$ regions of pixels.
The clip limit is sampled from the uniform distribution $\mathcal{U}(30, 50)$.

\subsection{Gamma Correction (U04)}

The pixel intensities of the image are nonlinearly modified.
Pixel intensity $I$ is transformed as follows:

\begin{equation}
    I' \leftarrow 255 \, \Big(\frac{I}{255}\Big)^\gamma
\end{equation}

where $\gamma \sim \mathcal{U}(0.5, 1.75)$.
The gain is fixed at $1$.

\subsection{Brightness and Contrast Change (U05)}

The brightness and contrast of the image are modified.
The brightness change factor, contrast change factor are sampled from $\mathcal{U}(0.6, 1.4)$, $\mathcal{U}(0.6, 1.4)$, respectively.
The image is then multiplied element-wise by its FOV mask, to ensure black regions external to the FOV remain black.

\subsection{Depth Change Simulation (U06)}

preserves the centre for linear FOV shapes and preserves $p_0$ for convex FOV shapes.
The magnitude of the zoom transformation, $d$, is randomly sampled from a uniform distribution.
Increasing the depth corresponds to zooming out ($d > 1$), while decreasing the depth corresponds to zooming in ($d < 1$).

\subsection{Speckle Noise Simulation (U07)}

Following Singh~\etal's method~\cite{singh2017synthetic}, this transformation calculates synthetic speckle noise and applies it to the ultrasound FOV. 
Various parameters of the algorithm are randomly sampled upon each invocation.
The lateral and axial resolutions for interpolation are random integers drawn from the ranges $[35, 45]$ and $[75, 85]$, respectively.
The number of synthetic phasors is randomly drawn from the integer range $[5, 10]$.
Sample points on the image are evenly spaced in Cartesian coordinates for linear FOV shapes.
For convex FOVs, the sample points are evenly spaced in polar coordinates.

\subsection{Gaussian Noise Simulation (U08)}

Multiplicative Gaussian noise is applied to the pixel intensities across the image.
First, the standard deviation of the Gaussian noise, $\sigma$, is randomly drawn from the uniform distribution $\mathcal{U}(0.5, 2.5)$.
Multiplicative Gaussian noise with mean $1$ and standard deviation $\sigma$ is then applied independently to each pixel in the image.

\subsection{Salt and Pepper Noise Simulation (U09)}

A random assortment of points in the image are set to $255$ (salt) or $0$ (pepper).
The fractions of pixels set to salt and pepper values are sampled randomly according to $\mathcal{U}(0.001, 0.005)$.

\subsection{Horizontal Reflection (U10)}

The image is reflected about the central vertical axis.
This transformation is identical to U01.

\subsection{Rotation and Shift (U11)}

A non-scaling affine transformation is applied to the image.
More specifically, the image is translated and rotated.
The horizontal component of the translation is sampled from $\mathcal{U}(-0.2, 0.2)$, as a fraction of the image's width.
Similarly, the vertical component is sampled from $\mathcal{U}(-0.2, 0.2)$, as a fraction of the image's height.
The rotation angle, in degrees, is sampled from $\mathcal{U}(-22.5, 22.5)$.

\section{Pleural Line Object Detection Training}
\label{apx:pl-od-training}

The Single Shot Detector (SSD) method~\cite{liu2016ssd} was employed to perform a cursory evaluation of the pretrained models on an object detection task.
The {\tt PL} task involved localizing instances of the pleural line in lung US images, which can be described as a bright horizontal line that is typically situated slightly below the vertical level of the ribs.
It is only visible between the rib spaces, since bone blocks the ultrasound scan lines.
The artifact represents the interface between the parietal and visceral linings of the lung.
In a properly acquired lung US view,

As in the classification experiments, we used the MobileNetV3Small architecture as the backbone of the network.
There is precedent for using the SSD object detection method, as it has been applied to assess the object detection capabilities of MobileNet architectures~\cite{sandler2018mobilenetv2,howard2019searching}. 
The feature maps outputted from a designated set of layers were passed to the SSD regression and classification heads.
We selected a range of layers whose feature maps had varying spatial resolution.
Table~\ref{tab:ssd-feat-maps} provides the identities and dimensions of the feature maps from the backbone that were fed to the SSD model head.
The head contained \num{43796} trainable parameters, which was light compared to the backbone (\num{927008}).

The set of default anchor box aspect ratios was manually specified after examining the distribution of bounding box aspect ratios in the training set.
The 25\textsuperscript{th} percentile was $2.894$, and the 75\textsuperscript{th} percentile was $4.989$.
The pleural line typically has a much greater width than height.
Accordingly, we designated the set of default anchor box aspect ratios ($w / h$) as $\{1, 2, 3, 4, 5\}$.
Six anchor box scales were used.
The first five were spaced out evenly over the range $[0.023, 0.170]$, which correspond to the square roots of the minimum and maximum areas of the bounding box labels present in the training set, in normalized image coordinates.
The final scale is $1.000$, which is included by default.
The box confidence threshold was $0.01$ and the intersection over union threshold to match anchors to box labels was $0.3$.
The non-maximum suppression (NMS) threshold was $0.45$ and the number of detected boxes to keep after NMS was $50$.

The backbone and head were assigned initial learning rates of $0.002$ and $0.02$, respectively.
Learning rates were annealed according to a cosine decay schedule.
The model was trained for $30$ epochs to minimize the loss function from the original SSD paper, which has a regression component for bounding box offsets and a classification component for distinguishing objects and background~\cite{liu2016ssd}.
The weights corresponding to the epoch with the lowest validation loss were retained for test set evaluation.

\begin{table}[h!]
    \caption{MobileNetV3Small block indices and the corresponding dimensions of the feature maps that they output, given an input of size $128\times128\times3$.}
    \centering
    \begin{tabular}{c@{\hskip 1.3cm}c}
        \toprule
         Block index & Feature map dimensions ($w \times h \times c$) \\
         \midrule
         $1$ & $32 \times 32 \times 16$ \\
         $3$ & $16 \times 16 \times 24$ \\
         $6$ & $8 \times 8 \times 40$ \\
         $9$ & $4 \times 4 \times 96$ \\
         $12$ & $4 \times 4 \times 576$ \\
         \bottomrule
    \end{tabular}
    \label{tab:ssd-feat-maps}
\end{table}

\section{Leave-one-out Analysis Statistical Testing}
\label{apx:looa-stats}

As outlined in Section IV-A, statistical testing was performed to detect differences between pretrained models trained using an ablated version of the StandardAug and AugUS-O pipelines and baseline models pretrained on the original pipelines.
Each ablated version of the pipeline was missing one transformation from the data augmentation pipeline.
Ten-fold cross-validation conducted on the training set provided $10$ samples of test AUC metrics for both the A-line versus B-line ({\tt AB}) and pleural effusion ({\tt PE}) binary classification tasks.
The samples were taken as a proxy for test-time performance for linear classifiers trained on each of the above downstream tasks.

To determine whether the mean test AUC for each ablated model was different from the baseline model, hypothesis testing was conducted.
The model pretrained using the original pipeline was the control group, while the models pretrained using ablated versions of the pipeline were the experimental groups.
First, Friedman's test~\cite{friedman1940comparison} was conducted to determine if there was any difference in the mean test AUC among the baseline and ablated models.
We selected a nonparametric multiple comparisons test because of the lack of assumptions regarding normality or homogeneity of variances. 
Each collection had $10$ samples.
Table~\ref{tab:friedman} details the results of Friedman's test for each pipeline and classification task.
Friedman's test detected differences among the collection of test AUC for both classification tasks with the StandardAug pipeline.
Only the {\tt AB} task exhibited significant differences for the AugUS-O pipeline.

\begin{table}[h!]
    \centering
    \caption{Friedman test statistics and \textit{p}-values for mean cross-validation test AUC attained by models pretrained using an entire data augmentation pipeline and ablated versions of it.
    }
    \begin{threeparttable}
    \begin{tabular}{ccccc}
        \toprule
        Pipeline &  \multicolumn{2}{c}{\tt AB} & \multicolumn{2}{c}{\tt PE} \\
        & $F_r$ Statistic & \textit{p}-value  & $F_r$ Statistic & \textit{p}-value \\
        \midrule
        StandardAug & $30.73$ & $0.000$\tnote{*} & $36.51$ & $0.000$\tnote{*} \\
        AugUS-O &  $76.43$ & $0.000$\tnote{*} & $18.91$ & $0.091$ \\
        \bottomrule
    \end{tabular}
    \begin{tablenotes}
    \footnotesize
    \item[*] Statistically significant at $\alpha=0.05$.
    \end{tablenotes}
    \end{threeparttable}
    \label{tab:friedman}
\end{table}

When the null hypothesis of the Friedman test was rejected, post-hoc tests were conducted to determine whether any of the test AUC means in the experimental groups were significantly different than the control group.
The Wilcoxon Sign-Rank Test~\cite{wilcoxon1945} was designated as the post-hoc test, due to its absence of any normality assumptions.
Note that for each pipeline, $n$ comparisons were performed, where $n$ is the number of transformations within the pipeline.
The Holm-Bonferroni correction~\cite{holm1979simple} was applied to keep the family-wise error rate at $\alpha=0.05$ for each pipeline/task combination.
Results of the post-hoc tests are given in Table~\ref{tab:wilcoxon}.
No post-hoc tests were performed for the AugUS-O pipeline evaluated on the {\tt PE} task because the Friedman test revealed no significant differences.

\begin{table}[h!]
    \centering
    \caption{Test statistics ($T$) and \textit{p}-values obtained from the Wilcoxon Sign-Rank post-hoc tests that compared linear classifiers trained with ablated models' features to a control linear classifier trained on the baseline model.
    Experimental groups are identified according to the left-out transformation, as defined in Tables~II~and~III.
    }
    \footnotesize
    \begin{threeparttable}
    \begin{tabular}{cccccc}
        \toprule
        Pipeline & Comparison & \multicolumn{2}{c}{\tt AB} & \multicolumn{2}{c}{\tt PE} \\
        & &  $T$ & \textit{p}-value & $T$ & \textit{p}-value \\
        \midrule
        \multirow{6}{*}{{\small StandardAug}} &  {\it B00} & $0$  & $0.002$\tnote{*} & $0$ & $0.002$\tnote{*}\\
        & {\it B01} &  $6$ & $0.027$ & $21$ & $0.557$\\
        & {\it B02} & $1$  & $0.004$\tnote{*} & $3$ & $0.010$\tnote{*}\\
        & {\it B03} & $19$  & $0.432$ & $10$ & $0.084$\\
        & {\it B04} &  $9$ & $0.064$ & $5$ & $0.020$\\
        & {\it B05} & $15$  & $0.232$ & $10$ & $0.084$\\
        \midrule
        \multirow{12}{*}{\small AugUS-O} & {\it U00} & $18$ & $0.375$ & - & - \\
        & {\it U01} & $8$ & $0.049$ & -& -\\
        & {\it U02} &  $0$ & $0.002$\tnote{*} & -& -\\
        & {\it U03}  & $0$  & $0.002$\tnote{*} &- & -\\
        & {\it U04} &  $12$ & $0.131$ & -& -\\
        & {\it U05} &  $9$ & $0.064$ &- & -\\
        & {\it U06} &  $13$ & $0.160$ &- & -\\
        & {\it U07} &  $13$ & $0.160$ &- & -\\
        & {\it U08} &  $1$ & $0.004$\tnote{*} & -& -\\
        & {\it U09}  &  $1$ & $0.004$\tnote{*} &- & -\\
        & {\it U10} &  $23$ & $0.695$ &- & -\\
        & {\it U11}  &  $0$ & $0.002$\tnote{*} &- & -\\
        \bottomrule
    \end{tabular}
    \begin{tablenotes}
    \footnotesize
    \item[*] Statistically significant at family-wise error rate of $0.05$.
    \end{tablenotes}
    \end{threeparttable}
    \label{tab:wilcoxon}
\end{table}

\section{Additional Positive Pair Examples}
\label{apx:additional-positive-pairs}

\begin{figure*}
    \centerline{\includegraphics[width=\linewidth]{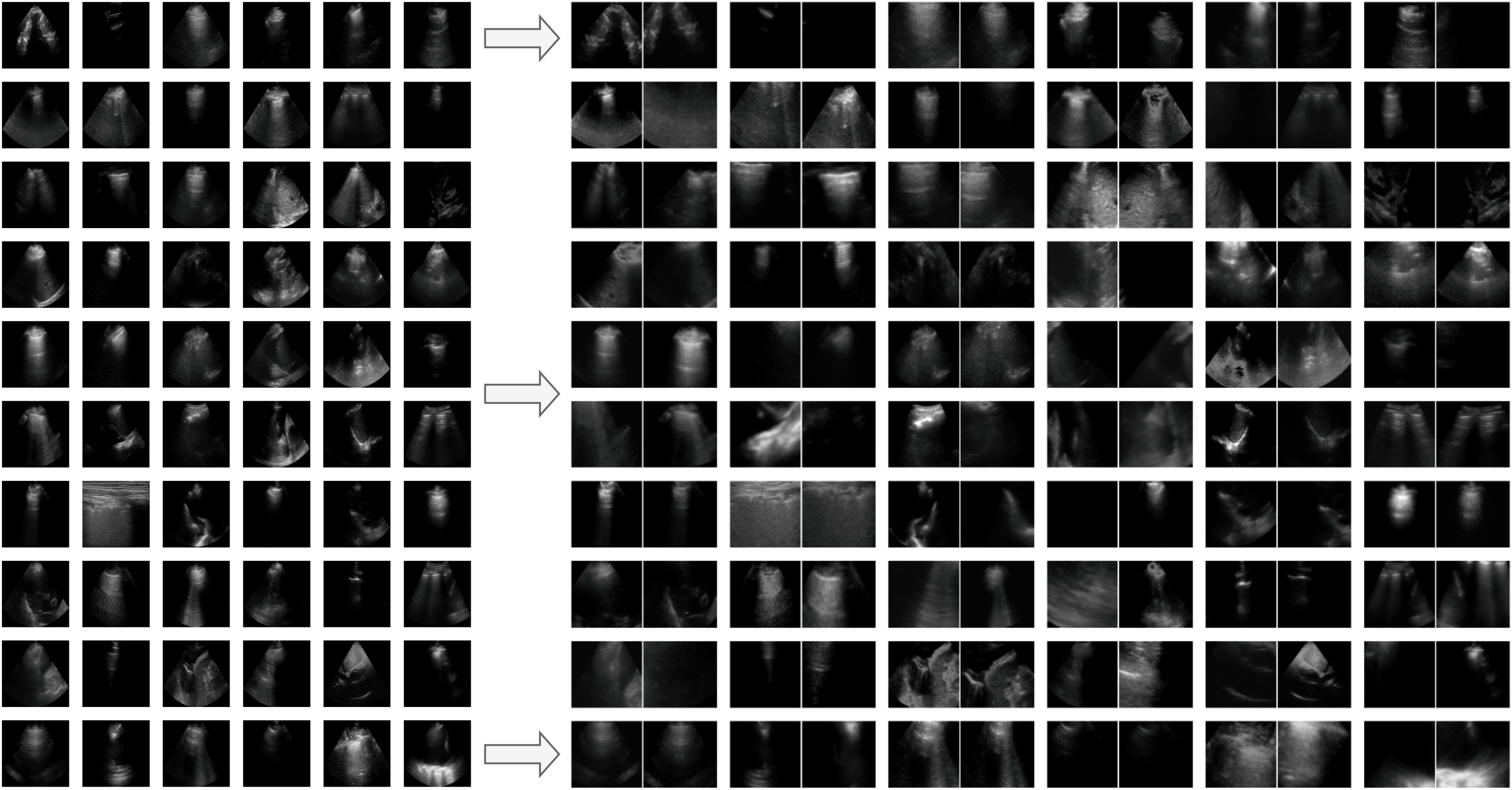}}
    \caption{Examples of lung ultrasound images (left) and positive pairs produced using the StandardAug pipeline (right).}
    \label{fig:byol-extra-examples}
\end{figure*}

\begin{figure*}
    \centerline{\includegraphics[width=\linewidth]{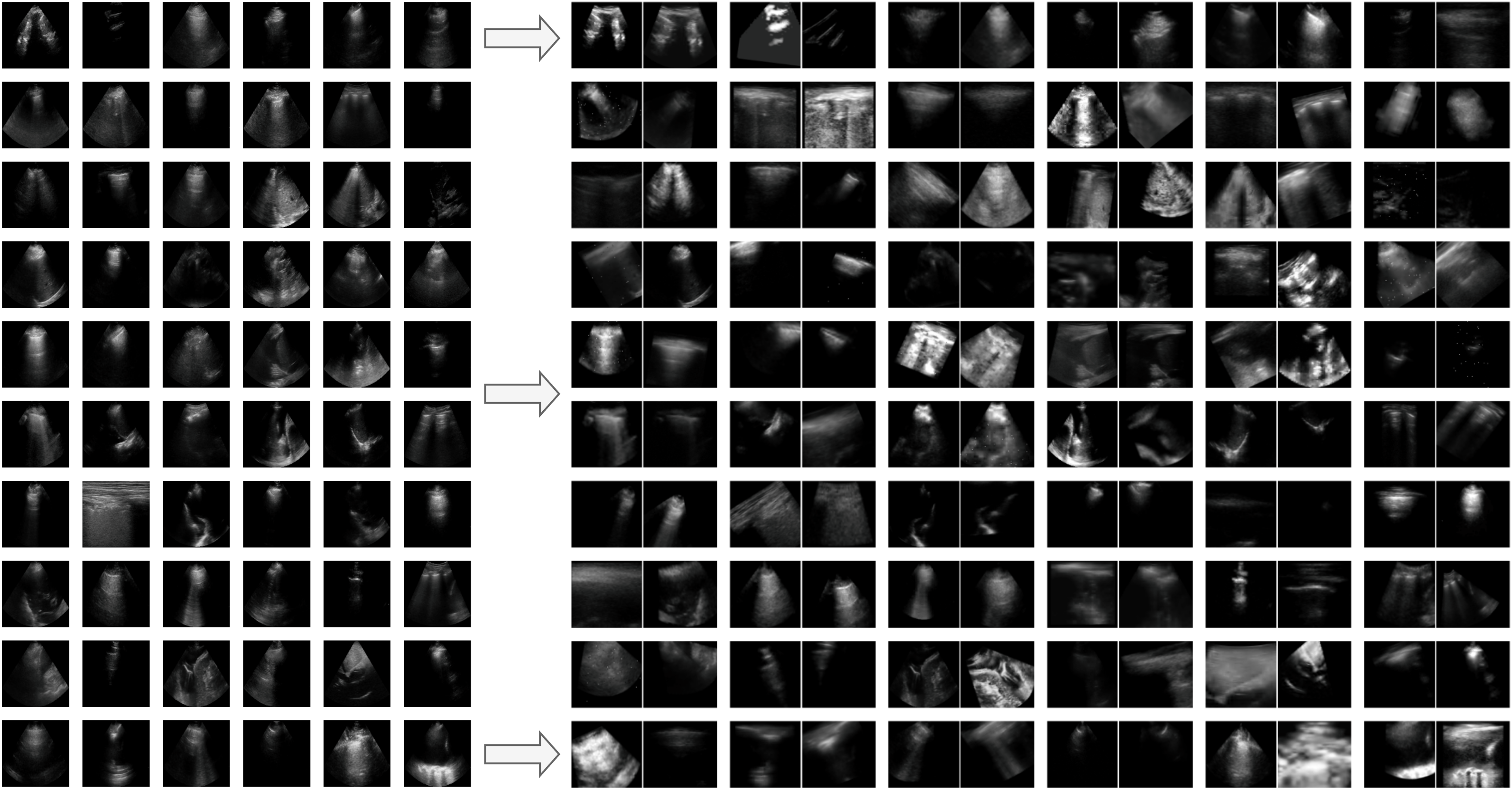}}
    \caption{Examples of lung ultrasound images (left) and positive pairs produced using the AugUS-O pipeline (right).}
    \label{fig:auguso-extra-examples}
\end{figure*}

\begin{figure*}
    \centerline{\includegraphics[width=\linewidth]{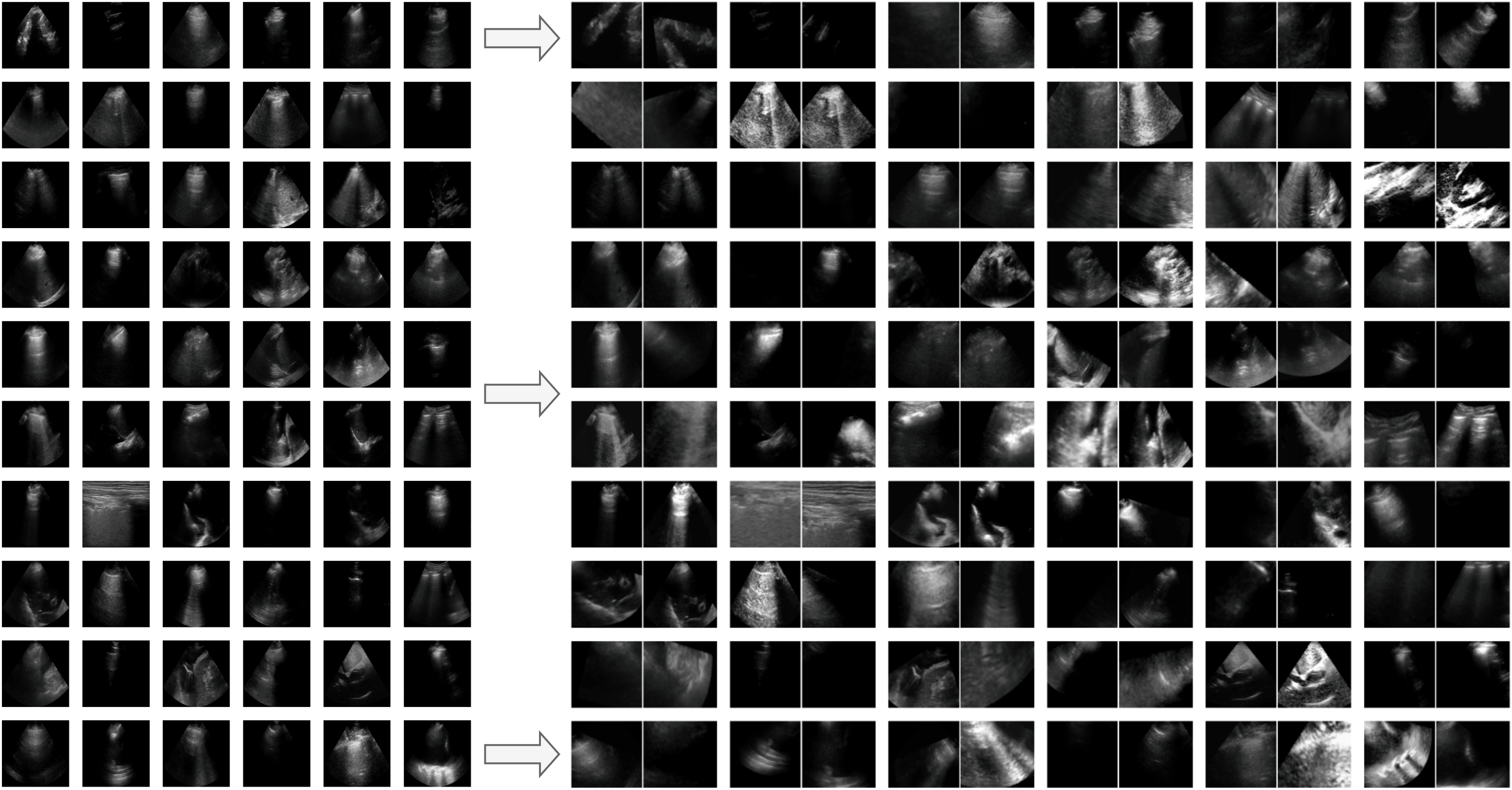}}
    \caption{Examples of lung ultrasound images (left) and positive pairs produced using the AugUS-D pipeline (right).}
    \label{fig:augusd-extra-examples}
\end{figure*}

Figs.~\ref{fig:byol-extra-examples},~\ref{fig:auguso-extra-examples},~and~\ref{fig:augusd-extra-examples} provide several examples of positive pairs produced by the StandardAug, AugUS-O, and AugUS-D pipelines, respectively.
Each figure shows original images from LUSData, along with two views of each image that were produced by applying stochastic data augmentation twice to the original images.

\section{Results with ResNet18 Backbone}
\label{apx:resnet18-experiments}

As outlined in Section~\ref{subsec:training-details}, MobileNetV3Small~\cite{howard2019searching} was selected as the feature extractor for all experiments in this study, primarily due to its suitability for lightweight inference in edge deployment scenarios.
It has also been used in prior work for similar tasks~\cite{vanberlo2023self,vanberlo2024intra}.
However, we wanted to determine if similar trends in the results held for an alternate backbone architecture with greater capacity.
Larger CNN backbones are frequently used for self-supervised pretraining.

We repeated the experiments in Sections~\ref{subsec:object-classification}~and~\ref{subsec:diagnostic-classification-task-evaluation} using ResNet18~\cite{he2016deep} as the feature extractor, which is a more common architecture.
Note that ResNet18's capacity is far greater than that of MobileNetV3Small --- the former has approximately \num{11200000} trainable parameters, while the latter has about \num{927000}.
We applied the same architecture for the projector head that was used for MobileNetV3Small experiments, which was a $2$-layer multilayer perceptron with $576$ nodes comprising each layer. 

Table~\ref{tab:test-results-internal-resnet18} displays the performance of pretrained ResNet18 backbones when evaluated on the LUSData test set for the {\tt AB}, {\tt PE}, and {\tt COVID} tasks.
Most fine-tuned models exhibited overfitting, likely due to the architecture's capacity.
Notably, the pretrained models tended to exhibit overfitting when fine-tuned, while the fully supervised ImageNet-pretrained model achieved strong performance on {\tt AB} and {\tt PE}.
Linear classifiers trained on frozen pretrained backbones' output features generally achieved the strongest performance and avoided overfitting.
Consistent with MobileNetV3Small, the use of cropping-based augmentation pipelines translated to greater performance on {\tt AB} and {\tt PE} in the linear evaluation setting. 
In the fine-tuning setting, the backbone pretrained using AugUS-O exhibited markedly poorer performance on {\tt AB} but achieved the greatest test AUC on {\tt PE}.
On the {\tt COVID} test set, the backbone pretrained using the AugUS-O pipeline led to the greatest average AUC in both the linear and fine-tuning settings.

The fine-tuned backbones and linear classifiers were also evaluated on the external test set for the {\tt AB} and {\tt PE} tasks.
Again, linear classifiers with pretrained backbones mostly achieved greater performance than fine-tuned classifiers.
Backbones pretrained with the StandardAug pipeline achieved the greatest external test set AUC for the {\tt AB} task.
However, those pretrained with AugUS-O performed comparably with StandardAug for the {\tt PE} task under the linear evaluation setting but achieved the greatest AUC in the fine-tuning setting. 

Two key observations were drawn from these experiments. 
First, the low-capacity MobileNetV3Small backbones achieved similar performance to the high-capacity ResNet18 backbones for these LUS tasks when their weights were frozen (i.e., linear classifiers).
Second, with a higher-capacity backbone, linear classifiers trained on the features outputted by SSL-pretrained backbones often achieved greater performance than fine-tuning using despite the weight initialization strategy.
The opposite trend was observed for backbones initialized with ImageNet-pretrained weights.

\begin{table}[H]
    \centering
    \setlength{\tabcolsep}{0.13cm}
    \caption{Test set performance for linear classification (LC) and fine-tuning (FT) experiments with the {\tt COVID} task using pretrained ResNet18 backbones.
    For {\tt COVID}, metrics are averages across all four classes.
    The best observed metrics in each experimental setting are in \textbf{boldface}.}
    {\small
    \begin{tabular}{cccccccc}
        \toprule
         \makecell{Train \\ Setting} & Task & Weights & Pipeline & Accuracy & Precision & Recall & AUC \\
         \midrule
          \multirow{12}{*}{{\large LC}} & \multirow{4}{*}{{\tt AB}} & SimCLR & StandardAug & $\mathbf{0.932}$ & $0.929$ & $\mathbf{0.845}$  & $\mathbf{0.968}$\\
          & & SimCLR & AugUS-O & $0.916$ & $0.923$ & $0.792$ &  $0.954$\\
          & &  SimCLR & AugUS-D & $0.931$ & $\mathbf{0.951}$ & $0.819$ &  $0.966$\\
          & & ImageNet & - & $0.878$ & $0.838$ & $0.749$ & $0.933$ \\
            \cmidrule(l{3pt}){2-8} 
          & \multirow{4}{*}{{\tt PE}} & SimCLR & StandardAug & $0.781$ & $0.772$ & $0.777$ & $\mathbf{0.879}$\\
          & & SimCLR & AugUS-O & $\mathbf{0.794}$ & $\mathbf{0.806}$ & $0.759$ &  $0.878$\\
          & & SimCLR & AugUS-D & $0.779$ & $0.739$ & $\mathbf{0.840}$ &  $0.873$\\
          & & ImageNet & - & $0.732$ & $0.693$ & $0.799$ & $0.823$ \\
           \cmidrule(l{3pt}){2-8} 
          & \multirow{4}{*}{{\tt COVID}} & SimCLR & StandardAug & $0.537$ & $0.538$ & $0.496$ & $\mathbf{0.827}$\\
          & & SimCLR & AugUS-O & $\mathbf{0.567}$ & $\mathbf{0.438}$ & $\mathbf{0.515}$ &  $0.782$\\
          & & SimCLR & AugUS-D & $0.487$ & $0.353$ & $0.447$ & $0.789$\\
          & & ImageNet & - & $0.527$ & $0.334$ & $0.476$ & $0.699$ \\
         \midrule
         
          \multirow{15}{*}{{\large FT}} & \multirow{5}{*}{{\tt AB}} & SimCLR & StandardAug & $0.882$ & $0.796$ & $0.824$  & $0.929$ \\
          & & SimCLR & AugUS-O & $0.375$ & $0.328$ & $\mathbf{0.990}$ &  $0.867$\\
          & & SimCLR & AugUS-D & $0.498$ & $0.377$ & $0.979$ &  $0.907$\\
          & & Random & - & $\mathbf{0.919}$ & $\mathbf{0.898}$ & $0.830$ & $0.960$  \\
          & & ImageNet & - & $0.892$ & $0.778$ & $0.906$ & $\mathbf{0.961}$ \\
          \cmidrule(l{3pt}){2-8}
          & \multirow{5}{*}{{\tt PE}} & SimCLR & StandardAug & $0.692$ & $0.643$ & $0.817$ & $0.795$\\
          & & SimCLR & AugUS-O & $0.773$ & $\mathbf{0.928}$ & $0.576$ &  $\mathbf{0.887}$\\
          & & SimCLR & AugUS-D & $0.654$ & $0.601$ & $\mathbf{0.852}$ &  $0.755$\\   
          & & Random & - & $0.720$ & $0.684$ & $0.784$ & $0.815$  \\
          & & ImageNet & - & $\mathbf{0.778}$ & $0.797$ & $0.727$ & $0.852$  \\
          \cmidrule(l{3pt}){2-8} 
          & \multirow{4}{*}{{\tt COVID}} & SimCLR & StandardAug & $0.418$ & $\mathbf{0.451}$ & $0.447$ & $0.719$\\
          & & SimCLR & AugUS-O & $\mathbf{0.564}$ & $0.434$ & $\mathbf{0.511}$ &  $\mathbf{0.723}$\\
          & & SimCLR & AugUS-D & $0.431$ & $0.346$ & $0.376$ & $0.631$\\
          & & Random & - & $0.372$ & $0.277$ & $0.354$ & $0.629$ \\
          & & ImageNet & - & $0.549$ & $0.433$ & $0.496$ & $0.719$ \\
         \bottomrule
        
    \end{tabular}}
    \label{tab:test-results-internal-resnet18}
\end{table}

\begin{table}[H]
    \centering
    \setlength{\tabcolsep}{0.14cm}
    \caption{External test set performance for linear classifiers (LC) and fine-tuned models (FT). 
    The best observed metrics in each experimental setting are in \textbf{boldface}.}
    {\small
    \begin{tabular}{cccccccc}
        \toprule
          \makecell{Train \\ Setting} & Task & \makecell{Initial \\ Weights} & Pipeline & Accuracy & Precision & Recall & AUC \\

         \midrule 
          \multirow{8}{*}{{\large LC}} & \multirow{5}{*}{{\tt \large AB}} & SimCLR & StandardAug & $\mathbf{0.762}$ & $\mathbf{0.956}$ & $\mathbf{0.581}$ & $\mathbf{0.877}$ \\
          & &  SimCLR & AugUS-O & $0.713$ & $0.939$ & $0.494$ & $0.826$ \\
          & & SimCLR & AugUS-D & $0.739$ & $0.945$ & $0.542$ & $0.862$ \\
          & & ImageNet & - & $0.659$ & $0.904$ & $0.402$ & $0.785$ \\
          \cmidrule(l{3pt}){2-8}
          & \multirow{5}{*}{{\tt \large PE}} & SimCLR & StandardAug & $\mathbf{0.797}$ & $0.884$ & $0.784$ & $\mathbf{0.887}$ \\
          & & SimCLR & AugUS-O & $0.786$ & $\mathbf{0.911}$ & $0.737$ & $\mathbf{0.887}$ \\
          & & SimCLR & AugUS-D & $0.772$ & $0.818$ & $\mathbf{0.826}$ & $0.851$ \\
          & & ImageNet & - & $0.701$ & $0.790$ & $0.723$ & $0.776$ \\
          \midrule
          
          \multirow{10}{*}{{\large FT}} & \multirow{5}{*}{{\tt \large AB}} & SimCLR & StandardAug & $0.767$ & $\mathbf{0.951}$ & $0.594$ & $\mathbf{0.880}$ \\ 
          & & SimCLR & AugUS-O & $0.597$ & $0.576$ & $\mathbf{0.924}$ & $0.728$ \\
          & & SimCLR & AugUS-D & $0.697$ & $0.666$ & $0.867$ & $0.816$ \\
          & & Random & - & $0.744$ & $0.919$ & $0.576$ & $0.851$ \\
          & & ImageNet & - & $\mathbf{0.771}$ & $0.855$ & $0.688$ & $0.857$ \\
          \cmidrule(l{3pt}){2-8}
          & \multirow{5}{*}{{\tt \large PE}} & SimCLR & StandardAug & $0.612$ & $0.791$ & $0.532$ & $0.700$ \\
          & & SimCLR & AugUS-O & $0.678$ & $\mathbf{0.972}$ & $0.509$ & $\mathbf{0.897}$ \\
          & & SimCLR & AugUS-D & $0.703$ & $0.752$ & $\mathbf{0.797}$ & $0.755$ \\
          & & Random & - & $0.743$ & $0.845$ & $0.731$ & $0.809$ \\
          & & ImageNet & - & $\mathbf{0.764}$ & $0.897$ & $0.712$ & $0.854$ \\
         \bottomrule
    \end{tabular}
    }
    \label{tab:test-results-external-resnet18}
\end{table}

\section{Label Efficiency Statistical Testing}
\label{apx:label-eff-stats}

For each of the {\tt AB} and {\tt PE} tasks, there were five experimental conditions: SimCLR pretraining with the StandardAug pipeline, SimCLR pretraining with the AugUS-O pipeline, SimCLR pretraining with the AugUS-D pipeline, ImageNet weight initializations, and random weight initialization.
The population consisted of $20$ subsets of the training set that were split randomly by patient.
The same splits were used across all conditions, reflecting a repeated measures design.
The experiment was repeated separately for the {\tt AB} and the {\tt PE} task.

The Friedman Test Statistic ($F_r$) was $75.44$ for {\tt AB}, with a \textit{p}-value of $0$.
For {\tt PE}, the $F_r = 45.44$, and the \textit{p}-value was $0$.
As such, the null hypothesis was rejected for both cases, indicating the existence of differences among the mean test AUC across conditions.
The Wilcoxon Sign-Rank test was performed as a post-hoc test between each pair of populations.
Tables~\ref{tab:wilcoxon-label-eff-ab}~and~\ref{tab:wilcoxon-label-eff-pe} provide all Wilcoxon Test Statistics, along with $p$-values and differences of the medians between conditions.
The Bonferroni correction was applied to the $p$-values to keep the family-wise error rate to $\alpha=0.05$.
Statistically significant comparisons are indicated.

\begin{table}[h!]
    \centering
    \caption{Test statistics ($T$) and \textit{p}-values obtained from the Wilcoxon Sign-Rank post-hoc tests comparing LUSData test AUC on {\tt AB} for classifiers trained on subsets of the training set.
    For comparison $a / b$, $\Delta := \text{median}(b) - \text{median}(a)$.
    The displayed $p-$values have been adjusted according to the Bonferroni correction to control the family-wise error rate.
    }
    \begin{threeparttable}
    \begin{tabular}{lccc}
        \toprule
        Comparison & $T$ &  \textit{p}-value & $\Delta$ \\
        \midrule
           Random / ImageNet & $3$   &  \num{9.4e-5}\tnote{*} & $-0.058$ \\
           Random / StandardAug & $0$ & \num{1.9e-5}\tnote{*} & $0.044$\\
           Random / AugUS-O & $12$  & \num{1.3e-3}\tnote{*} & $0.019$\\
           Random / AugUS-D & $0$  & \num{1.9e-5}\tnote{*} & $0.042$ \\
           ImageNet / StandardAug & $0$  & \num{1.9e-5}\tnote{*} & $0.100$ \\
           ImageNet / AugUS-O & $0$ & \num{1.9e-5}\tnote{*} & $0.078$ \\
           ImageNet / AugUS-D & $0$ & \num{1.9e-5}\tnote{*} & $0.100$ \\
           StandardAug / AugUS-O & $0$  & \num{1.9e-5}\tnote{*} & $-0.024$ \\
           StandardAug / AugUS-D & $61$  & \num{1.0e-0} & $-0.002$\\
           AugUS-O / AugUS-D & $0$  & \num{1.9e-5}\tnote{*} & $0.022$\\
        \bottomrule
    \end{tabular}
    \begin{tablenotes}
    \footnotesize
    \item[*] Statistically significant at family-wise error rate of $0.05$.
    \end{tablenotes}
    \end{threeparttable}
    \label{tab:wilcoxon-label-eff-ab}
\end{table}

\begin{table}[h!]
    \centering
    \caption{Test statistics ($T$) and \textit{p}-values obtained from the Wilcoxon Sign-Rank post-hoc tests comparing LUSData test AUC on {\tt PE} for classifiers trained on subsets of the training set.
    For comparison $a / b$, $\Delta := \text{median}(b) - \text{median}(a)$.
    The displayed $p-$values have been adjusted according to the Bonferroni correction to control the family-wise error rate.
    }
    \begin{threeparttable}
    \begin{tabular}{lccc}
        \toprule
        Comparison & $T$ &  \textit{p}-value & $\Delta$ \\
        \midrule
           Random / ImageNet & $1$   &  \num{3.8e-5}\tnote{*} & $-0.127$ \\
           Random / StandardAug & $31$ & \num{4.2e-2}\tnote{*} & $0.025$\\
           Random / AugUS-O & $16$  & \num{3.2e-3}\tnote{*} & $0.030$\\
           Random / AugUS-D & $5$  & \num{1.9e-4}\tnote{*} & $0.036$ \\
           ImageNet / StandardAug & $1$  & \num{3.8e-5}\tnote{*} & $0.152$ \\
           ImageNet / AugUS-O & $0$ & \num{1.9e-5}\tnote{*} & $0.157$ \\
           ImageNet / AugUS-D & $1$ & \num{3.8e-5}\tnote{*} & $0.163$ \\
           StandardAug / AugUS-O & $57$  & \num{7.6e-1} & $0.005$ \\
           StandardAug / AugUS-D & $53$  & \num{5.3e-1} & $0.011$\\
           AugUS-O / AugUS-D & $78$  & \num{1e-0} & $0.006$\\
        \bottomrule
    \end{tabular}
    \begin{tablenotes}
    \footnotesize
    \item[*] Statistically significant at family-wise error rate of $0.05$.
    \end{tablenotes}
    \end{threeparttable}
    \label{tab:wilcoxon-label-eff-pe}
\end{table}

\section{Additional Random Crop and Resize Experiments}
\label{apx:rcr-experiments}

The C\&R transform encourages pretrained representations to be invariant to scale.
It is also believed that the C\&R transform instills invariance between global and local views or between disjoint views of the same object type~\cite{chen2020simple}.
While the minimum area of the crop determines the magnitude of the scaling transformations, the aspect ratio range dictates the difference in distortion in both axes of the image.
The default aspect ratio range is $[0.75, 1.33]$.
We pretrained with the AugUS-D pipeline using a fixed aspect range of $1$ and $c=0.08$, which resulted in test AUC of $0.971$ for {\tt AB} and $0.881$ for {\tt PE} in the LC training setting. 
Compared to the regular AugUS-D that uses the default aspect ratio range (Table~\ref{tab:augus-internal-test-results}), {\tt AB} test AUC remained unchanged, and {\tt PE} test AUC decreased by $0.005$.

Lastly, we conducted pretraining on LUSData using only the C\&R transformation; that is, the data augmentation pipeline was \textit{[B00]}.
Recent work by Moutakanni~\etal~\cite{moutakanniyou2024} suggests that, with sufficient quantities of training data, competitive performance in downstream computer vision tasks can be achieved using crop and resize as the sole transformation in joint embedding SSL.
Linear evaluation of a feature extractor pretrained with only C\&R yielded test AUC of $0.964$ and $0.874$ on {\tt AB} and {\tt PE}, respectively.
Compared to the linear evaluations presented in Section~4, these metrics are greater than those achieved using AugUS-O, but less than those achieved with StandardAug or AugUS-D. 
It is evident that C\&R is a powerful transformation for detecting local objects.

\clearpage
\bibliographystyle{elsarticle-num}
\bibliography{references}

\begin{thebibliography}{10}
\expandafter\ifx\csname url\endcsname\relax
  \def\url#1{\texttt{#1}}\fi
\expandafter\ifx\csname urlprefix\endcsname\relax\def\urlprefix{URL }\fi
\expandafter\ifx\csname href\endcsname\relax
  \def\href#1#2{#2} \def\path#1{#1}\fi

\bibitem{wang2021deep}
Y.~Wang, X.~Ge, H.~Ma, S.~Qi, G.~Zhang, Y.~Yao, Deep learning in medical ultrasound image analysis: a review, IEEE Access 9 (2021) 54310--54324.

\bibitem{yang2020improving}
Q.~Yang, J.~Wei, X.~Hao, D.~Kong, X.~Yu, T.~Jiang, J.~Xi, W.~Cai, Y.~Luo, X.~Jing, et~al., Improving {B}-mode ultrasound diagnostic performance for focal liver lesions using deep learning: A multicentre study, EBioMedicine 56 (2020).

\bibitem{ghorbani2020deep}
A.~Ghorbani, D.~Ouyang, A.~Abid, B.~He, J.~H. Chen, R.~A. Harrington, D.~H. Liang, E.~A. Ashley, J.~Y. Zou, Deep learning interpretation of echocardiograms, NPJ digital medicine 3~(1) (2020) 10.

\bibitem{vanberlo2022accurate}
B.~VanBerlo, D.~Wu, B.~Li, M.~A. Rahman, G.~Hogg, B.~VanBerlo, J.~Tschirhart, A.~Ford, J.~Ho, J.~McCauley, et~al., Accurate assessment of the lung sliding artefact on lung ultrasonography using a deep learning approach, Computers in biology and medicine 148 (2022) 105953.

\bibitem{liu2019deep}
S.~Liu, Y.~Wang, X.~Yang, B.~Lei, L.~Liu, S.~X. Li, D.~Ni, T.~Wang, Deep learning in medical ultrasound analysis: a review, Engineering 5~(2) (2019) 261--275.

\bibitem{ansari2024advancements}
M.~Y. Ansari, I.~A.~C. Mangalote, P.~K. Meher, O.~Aboumarzouk, A.~Al-Ansari, O.~Halabi, S.~P. Dakua, Advancements in deep learning for {B}-mode ultrasound segmentation: A comprehensive review, IEEE Transactions on Emerging Topics in Computational Intelligence (2024).

\bibitem{vanberlo2024survey}
B.~VanBerlo, J.~Hoey, A.~Wong, A survey of the impact of self-supervised pretraining for diagnostic tasks in medical {X}-ray, {CT}, {MRI}, and ultrasound, BMC Medical Imaging 24~(1) (2024) 79.

\bibitem{Balestriero2022}
R.~Balestriero, Y.~LeCun, Contrastive and non-contrastive self-supervised learning recover global and local spectral embedding methods, Advances in Neural Information Processing Systems 35 (2022) 26671--26685.

\bibitem{azizi_big_2021}
S.~Azizi, B.~Mustafa, F.~Ryan, Z.~Beaver, J.~Freyberg, J.~Deaton, A.~Loh, A.~Karthikesalingam, S.~Kornblith, T.~Chen, et~al., Big self-supervised models advance medical image classification, in: Proceedings of the IEEE/CVF international conference on computer vision, 2021, pp. 3478--3488.

\bibitem{zhao2021longitudinal}
Q.~Zhao, Z.~Liu, E.~Adeli, K.~M. Pohl, Longitudinal self-supervised learning, Medical image analysis 71 (2021) 102051.

\bibitem{basu2022unsupervised}
S.~Basu, S.~Singla, M.~Gupta, P.~Rana, P.~Gupta, C.~Arora, Unsupervised contrastive learning of image representations from ultrasound videos with hard negative mining, in: International Conference on Medical Image Computing and Computer-Assisted Intervention, Springer, 2022, pp. 423--433.

\bibitem{cabannes2023ssl}
V.~Cabannes, B.~Kiani, R.~Balestriero, Y.~LeCun, A.~Bietti, The {SSL} interplay: Augmentations, inductive bias, and generalization, in: International Conference on Machine Learning, PMLR, 2023, pp. 3252--3298.

\bibitem{chen2020simple}
T.~Chen, S.~Kornblith, M.~Norouzi, G.~Hinton, A simple framework for contrastive learning of visual representations, in: International conference on machine learning, PMLR, 2020, pp. 1597--1607.

\bibitem{grill2020bootstrap}
J.-B. Grill, F.~Strub, F.~Altch{\'e}, C.~Tallec, P.~Richemond, E.~Buchatskaya, C.~Doersch, B.~Avila~Pires, Z.~Guo, M.~Gheshlaghi~Azar, et~al., Bootstrap your own latent-a new approach to self-supervised learning, Advances in neural information processing systems 33 (2020) 21271--21284.

\bibitem{zbontar2021barlow}
J.~Zbontar, L.~Jing, I.~Misra, Y.~LeCun, S.~Deny, Barlow twins: Self-supervised learning via redundancy reduction, in: International Conference on Machine Learning, 2021, pp. 12310--12320.

\bibitem{bardes2022vicreg}
A.~Bardes, J.~Ponce, Y.~LeCun, {VICR}eg: {Variance-Invariance-Covariance Regularization for Self-Supervised Learning}, in: International Conference on Learning Representations, 2022.

\bibitem{fernandez2022contrasting}
A.~Fernandez-Quilez, T.~Eftest{\o}l, S.~R. Kjosavik, M.~Goodwin, K.~Oppedal, Contrasting axial {T2W} mri for prostate cancer triage: A self-supervised learning approach, in: 2022 IEEE 19th International Symposium on Biomedical Imaging (ISBI), IEEE, 2022, pp. 1--5.

\bibitem{anand_benchmarking_2022}
D.~Anand, P.~Annangi, P.~Sudhakar, Benchmarking self-supervised representation learning from a million cardiac ultrasound images, in: Proceedings of the {Annual} {International} {Conference} of the {IEEE} {Engineering} in {Medicine} and {Biology} {Society}, {EMBS}, Vol. 2022-July, Institute of Electrical and Electronics Engineers Inc., 2022, pp. 529--532, iSSN: 1557170X.

\bibitem{saeed_contrastive_2022}
M.~Saeed, R.~Muhtaseb, M.~Yaqub, Contrastive pretraining for echocardiography segmentation with limited data, Lecture Notes in Computer Science (including subseries Lecture Notes in Artificial Intelligence and Lecture Notes in Bioinformatics) 13413 LNCS (2022) 680--691, iSBN: 9783031120527 Publisher: Springer Science and Business Media Deutschland GmbH.

\bibitem{nguyen2021semi}
N.-Q. Nguyen, T.-S. Le, A semi-supervised learning method to remedy the lack of labeled data, in: 2021 15th International Conference on Advanced Computing and Applications (ACOMP), IEEE, 2021, pp. 78--84.

\bibitem{chen2021uscl}
Y.~Chen, C.~Zhang, L.~Liu, C.~Feng, C.~Dong, Y.~Luo, X.~Wan, {USCL}: Pretraining deep ultrasound image diagnosis model through video contrastive representation learning, in: Medical Image Computing and Computer Assisted Intervention--MICCAI 2021: 24th International Conference, Strasbourg, France, September 27--October 1, 2021, Proceedings, Part VIII 24, Springer, 2021, pp. 627--637.

\bibitem{chen2022generating}
Y.~Chen, C.~Zhang, C.~H. Ding, L.~Liu, Generating and weighting semantically consistent sample pairs for ultrasound contrastive learning, IEEE Transactions on Medical Imaging (2022).

\bibitem{zhang2022hico}
C.~Zhang, Y.~Chen, L.~Liu, Q.~Liu, X.~Zhou, Hi{C}o: Hierarchical contrastive learning for ultrasound video model pretraining, in: Proceedings of the Asian Conference on Computer Vision, 2022, pp. 229--246.

\bibitem{vanberlo2024intra}
B.~VanBerlo, A.~Wong, J.~Hoey, R.~Arntfield, \href{https://www.frontiersin.org/journals/imaging/articles/10.3389/fimag.2024.1416114}{Intra-video positive pairs in self-supervised learning for ultrasound}, Frontiers in Imaging 3 (2024).
\newblock \href {https://doi.org/10.3389/fimag.2024.1416114} {\path{doi:10.3389/fimag.2024.1416114}}.
\newline\urlprefix\url{https://www.frontiersin.org/journals/imaging/articles/10.3389/fimag.2024.1416114}

\bibitem{chen2023contrastive}
L.~Chen, J.~Rubin, J.~Ouyang, N.~Balaraju, S.~Patil, C.~Mehanian, S.~Kulhare, R.~Millin, K.~W. Gregory, C.~R. Gregory, et~al., Contrastive self-supervised learning for spatio-temporal analysis of lung ultrasound videos, in: 2023 IEEE 20th International Symposium on Biomedical Imaging (ISBI), IEEE, 2023, pp. 1--5.

\bibitem{Ebadi2022-mn}
A.~Ebadi, P.~Xi, A.~MacLean, A.~Florea, S.~Tremblay, S.~Kohli, A.~Wong, {COVIDx-US}: An open-access benchmark dataset of ultrasound imaging data for {AI-driven} {COVID-19} analytics, Front. Biosci. (Landmark Ed.) 27~(7) (2022) 198.

\bibitem{zeng2024covid}
E.~Z. Zeng, A.~Ebadi, A.~Florea, A.~Wong, {COVID-Net L2C-ULTRA}: An explainable linear-convex ultrasound augmentation learning framework to improve {COVID-19} assessment and monitoring, Sensors 24~(5) (2024) 1664.

\bibitem{birge1997model}
L.~Birg{\'e}, P.~Massart, From model selection to adaptive estimation, in: D.~Pollard, E.~Torgersen, G.~L. Yang (Eds.), Festschrift for Lucien Le Cam: Research Papers in Probability and Statistics, Springer, New York, NY, 1997, pp. 55--87.
\newblock \href {https://doi.org/10.1007/978-1-4612-1880-7_4} {\path{doi:10.1007/978-1-4612-1880-7_4}}.

\bibitem{pizer1987}
S.~M. Pizer, E.~P. Amburn, J.~D. Austin, R.~Cromartie, A.~Geselowitz, T.~Greer, B.~ter Haar~Romeny, J.~B. Zimmerman, K.~Zuiderveld, Adaptive histogram equalization and its variations, Computer vision, graphics, and image processing 39~(3) (1987) 355--368.

\bibitem{vilimek2022comparative}
D.~Vilimek, J.~Kubicek, M.~Golian, R.~Jaros, R.~Kahankova, P.~Hanzlikova, D.~Barvik, A.~Krestanova, M.~Penhaker, M.~Cerny, et~al., Comparative analysis of wavelet transform filtering systems for noise reduction in ultrasound images, Plos one 17~(7) (2022) e0270745.

\bibitem{singh2017synthetic}
P.~Singh, R.~Mukundan, R.~de~Ryke, Synthetic models of ultrasound image formation for speckle noise simulation and analysis, in: 2017 International Conference on Signals and Systems (ICSigSys), IEEE, 2017, pp. 278--284.

\bibitem{howard2019searching}
A.~Howard, M.~Sandler, G.~Chu, L.-C. Chen, B.~Chen, M.~Tan, W.~Wang, Y.~Zhu, R.~Pang, V.~Vasudevan, et~al., {Searching for MobileNetV3}, in: Proceedings of the IEEE/CVF international conference on computer vision, 2019, pp. 1314--1324.

\bibitem{vanberlo2023self}
B.~VanBerlo, B.~Li, J.~Hoey, A.~Wong, Self-supervised pretraining improves performance and inference efficiency in multiple lung ultrasound interpretation tasks, IEEE Access 11 (2023) 135696--135707.

\bibitem{deng2009}
J.~Deng, W.~Dong, R.~Socher, L.-J. Li, K.~Li, L.~Fei-Fei, Image{N}et: A large-scale hierarchical image database, in: 2009 IEEE conference on computer vision and pattern recognition, Ieee, 2009, pp. 248--255.

\bibitem{you2019large}
Y.~You, J.~Li, S.~Reddi, J.~Hseu, S.~Kumar, S.~Bhojanapalli, X.~Song, J.~Demmel, K.~Keutzer, C.-J. Hsieh, Large batch optimization for deep learning: Training bert in 76 minutes, arXiv preprint arXiv:1904.00962 (2019).

\bibitem{liu2016ssd}
W.~Liu, D.~Anguelov, D.~Erhan, C.~Szegedy, S.~Reed, C.-Y. Fu, A.~C. Berg, {SSD}: Single shot multibox detector, in: Computer Vision--ECCV 2016: 14th European Conference, Amsterdam, The Netherlands, October 11--14, 2016, Proceedings, Part I 14, Springer, 2016, pp. 21--37.

\bibitem{friedman1940comparison}
M.~Friedman, A comparison of alternative tests of significance for the problem of m rankings, The annals of mathematical statistics 11~(1) (1940) 86--92.

\bibitem{wilcoxon1945}
F.~Wilcoxon, \href{http://www.jstor.org/stable/3001968}{Individual comparisons by ranking methods}, Biometrics Bulletin 1~(6) (1945) 80--83.
\newline\urlprefix\url{http://www.jstor.org/stable/3001968}

\bibitem{holm1979simple}
S.~Holm, A simple sequentially rejective multiple test procedure, Scandinavian journal of statistics (1979) 65--70.

\bibitem{he2016deep}
K.~He, X.~Zhang, S.~Ren, J.~Sun, Deep residual learning for image recognition, in: Proceedings of the IEEE conference on computer vision and pattern recognition, 2016, pp. 770--778.

\bibitem{blazic2023use}
I.~Blazic, C.~Cogliati, N.~Flor, G.~Frija, M.~Kawooya, M.~Umbrello, S.~Ali, M.-L. Baranne, Y.-J. Cho, R.~Pitcher, et~al., The use of lung ultrasound in {COVID-19}, ERJ Open Research 9~(1) (2023).

\bibitem{kim2023point}
K.~Kim, F.~Macruz, D.~Wu, C.~Bridge, S.~McKinney, A.~A. Al~Saud, E.~Sharaf, A.~Pely, P.~Danset, T.~Duffy, et~al., Point-of-care ai-assisted stepwise ultrasound pneumothorax diagnosis, Physics in Medicine \& Biology 68~(20) (2023) 205013.

\bibitem{sandler2018mobilenetv2}
M.~Sandler, A.~Howard, M.~Zhu, A.~Zhmoginov, L.-C. Chen, {MobileNetV2}: Inverted residuals and linear bottlenecks, in: Proceedings of the IEEE conference on computer vision and pattern recognition, 2018, pp. 4510--4520.

\bibitem{moutakanniyou2024}
T.~Moutakanni, M.~Oquab, M.~Szafraniec, M.~Vakalopoulou, P.~Bojanowski, You don’t need domain-specific data augmentations when scaling self-supervised learning, in: The Thirty-eighth Annual Conference on Neural Information Processing Systems.

\end{thebibliography}

\end{document}